\documentclass[11pt,a4paper]{article}
\pdfoutput=1

\usepackage{jheppub}
\usepackage{latexsym}
\usepackage{revsymb}
\usepackage{multirow}
\usepackage{color}
\usepackage[usenames,dvipsnames,svgnames,table]{xcolor}

\usepackage{graphicx}
\usepackage{epsfig}  
\usepackage{epsf}    
\usepackage{dcolumn}
\usepackage{bm}
\usepackage{dcolumn}
\usepackage{textcomp}
\usepackage{float}
\usepackage{hypcap}
\usepackage[]{hyperref}
\usepackage{adjustbox}
\usepackage{multirow}
\usepackage[utf8]{inputenc}
\usepackage[T1]{fontenc} 
\usepackage{subfigure}
\usepackage{alphalph}
\usepackage{morefloats}
\usepackage{textcomp}

\newcommand{\be}{\begin{equation}}
\newcommand{\ee}{\end{equation}}
\newcommand{\ba}{\begin{eqnarray}}
\newcommand{\ea}{\end{eqnarray}}

\preprint{IP/BBSR/2019-6, IPPP/19/74}

\title{Physics Potential of ESS$\nu$SB in the presence of a Light Sterile Neutrino} 

\author[a,b,c]{Sanjib Kumar Agarwalla,}
\author[d,a]{Sabya Sachi Chatterjee,}
\author[e,f]{Antonio Palazzo}

\affiliation[a]{Institute of Physics, Sachivalaya Marg, Sainik School Post, Bhubaneswar 751005, India}
\affiliation[b]{Homi Bhabha National Institute, Training School Complex, Anushakti Nagar, Mumbai 400085, India}
\affiliation[c]{International Centre for Theoretical Physics, Strada Costiera 11, Trieste 34151, Italy}
\affiliation[d]{Institute for Particle Physics Phenomenology, Department of Physics, Durham University, Durham, DH1 3LE, UK}
\affiliation[e]{Dipartimento Interateneo di Fisica ``Michelangelo Merlin'', Via Amendola 173, 70126 Bari, Italy} 
\affiliation[f]{Istituto Nazionale di Fisica Nucleare (INFN), Sezione di Bari, Via E.\ Orabona 4, I-70126 Bari, Italy}

\emailAdd{sanjib@iopb.res.in}
\emailAdd{sabya.s.chatterjee@durham.ac.uk}
\emailAdd{palazzo@ba.infn.it}

\abstract{ESS$\nu$SB is a proposed neutrino super-beam project at the ESS facility.
We study the performance of this setup in the presence of a light eV-scale sterile neutrino,
considering 540 km baseline with 2 years (8 years) of $\nu$  ($\bar\nu$) run-plan.
This baseline offers the possibility to work around the second oscillation maximum,
providing high sensitivity towards CP-violation (CPV). We explore in detail its capability in resolving CPV generated
by the standard CP phase $\delta_{13}$, the new CP phase $\delta_{14}$,
and the octant of $\theta_{23}$.  We find that 
the sensitivity to CPV induced by $\delta_{13}$ deteriorates noticeably 
when going from $3\nu$ to 4$\nu$ case. The two phases $\delta_{13}$ and $\delta_{14}$
can be reconstructed with a 1$\sigma$ uncertainty of
$\sim15^0$ and $	\sim35^0$ respectively. Concerning the octant of $\theta_{23}$, we find poor
sensitivity in both $3\nu$ and $4\nu$ schemes.
Our results show that a setup like ESS$\nu$SB working around the second oscillation maximum
with a baseline of 540 km, performs quite well to explore CPV in 3$\nu$ scheme, but it is not optimal for studying
CP properties in 3+1 scheme.}

\keywords{Neutrino Oscillation, Long-baseline, Sterile Neutrino, ESS$\nu$SB}
\arxivnumber{1909.13746}
\begin{document}
\maketitle

\section{Introduction and Motivation}
\label{introduction}

Several anomalous results recorded in short-baseline (SBL) experiments, 
indicate the existence of a fourth sterile neutrino (for reviews on this 
subject, see the references~\cite{Abazajian:2012ys,Palazzo:2013me,Gariazzo:2015rra,Giunti:2015wnd,Giunti:2019aiy,Boser:2019rta}) 
with mass $\sim$ 1\,eV.  The indications come from the accelerator experiments LSND~\cite{Aguilar:2001ty} and MiniBooNE~\cite{Aguilar-Arevalo:2018gpe}, and from the so-called reactor~\cite{Mention:2011rk} and Gallium~\cite{Hampel:1997fc,Abdurashitov:2005tb} anomalies. Constraints on light sterile neutrinos have been
derived also by the long-baseline (LBL) experiments MINOS and MINOS+~\cite{MINOS:2016viw,Adamson:2017uda}, 
NO$\nu$A~\cite{Adamson:2017zcg} and T2K~\cite{Abe:2019fyx}, by the reactor experiments Daya Bay~\cite{An:2016luf},%
\footnote{We also mention the work~\cite{Adamson:2016jku}, where the combination of MINOS, Daya-Bay, and Bugey-3 
was considered.}
DANSS~\cite{Danilov:2019aef} and NEOS~\cite{Ko:2016owz}, by the atmospheric neutrino data collected in Super-Kamiokande~\cite{Abe:2014gda}, IceCube~\cite{Aartsen:2017bap} and ANTARES~\cite{Albert:2018mnz}, and
by solar neutrinos~\cite{Giunti:2009xz,Palazzo:2011rj,Palazzo:2012yf}.

New SBL experiments are under construction, with the aim
of testing this intriguing hypothesis (see the review in~\cite{Lasserre:2014ita}).
The new SBL experiments are sensitive to the characteristic $L/E$ 
dependency  due to the oscillations intervening at the new mass-squared splitting.
This will allow them to measure with precision the value $\Delta m^2_{new} \sim 1$\,eV$^2$
and the new mixing angles of the sterile sector.
However, as already stressed in the literature, the SBL experiments will be unable
to furnish any information about the CP-violation (CPV) structure of the sterile sector.
Even the simplest extended framework involving only one neutrino state, the
so-called 3+1 scheme, entails two additional CP phases with respect to the
standard framework. Therefore, after a hypothetical discovery made at the SBL experiment,
we will face the problem of finding a way to determine these new CP phases.

In order to measure any CP phase one must be sensitive to the quantum interference of two different oscillation frequencies. 
In the 3+1 scenario, in SBL experiments, only the new frequency is observable, while both the standard
(solar and atmospheric) frequencies have no effect at all. For this reason the  SBL setups have no sensitivity
to CPV (both in 3-flavor and 3+1 schemes). As first shown in~\cite{Klop:2014ima}, 
things are qualitatively different in LBL setups, since in these experiments
the interference between two different frequencies becomes observable. In fact,
the LBL experiments are able to detect both the effects of the standard CP phase and those 
of the new ones. For this reason, the LBL experiments are complementary to the SBL ones
in nailing down the properties of sterile neutrinos.  

The new-generation of LBL experiments~\cite{Feldman:2012qt,Pascoli:2013wca,Agarwalla:2013hma,Agarwalla:2014fva,Abe:2015zbg,Stanco:2015ejj,Acciarri:2015uup,Abe:2016ero,Agarwalla:2017nld,Abi:2018dnh} 
are designed to have a central role in the search of CPV phenomena.
In this paper, we focus on the proposed super beam experiment to be performed at the 
European Spallation Source (ESS$\nu$SB). This facility will have a very powerful
neutrino beam with an average power of 5 MW, and the flux is expected to
peak around 0.25 GeV. We assume that these neutrinos will travel a distance of 540 km
providing the opportunity to work around the second oscillation maximum.

Our present study is complementary to other recent investigations performed 
about DUNE~\cite{Hollander:2014iha,Berryman:2015nua,Gandhi:2015xza,Agarwalla:2016xxa,Agarwalla:2016xlg,Coloma:2017ptb,Choubey:2017cba}, 
T2HK~\cite{Choubey:2017cba,Choubey:2017ppj,Agarwalla:2018t2hk}, and 
T2HKK~\cite{Choubey:2017cba,Haba:2018klh}.
Other studies on the impact of light sterile neutrinos in LBL setups can be found 
in~\cite{Donini:2001xy,Donini:2001xp,Donini:2007yf,Dighe:2007uf,Donini:2008wz,Yasuda:2010rj,Meloni:2010zr,Bhattacharya:2011ee,Donini:2012tt}.
We underline that while our study deals with charged current interactions, one can obtain valuable information on active-sterile oscillations parameters 
also from the analysis of neutral current interactions (see~\cite{MINOS:2016viw,Adamson:2017uda,Adamson:2017zcg,Abe:2019fyx} for constraints
from existing data  and~\cite{Coloma:2017ptb,Gandhi:2017vzo} for sensitivity studies of future experiments.)
 
The paper has the following structure. In section~\ref{sec:probability}, we detail the theoretical
framework and also describe the properties of the 4-flavor $\nu_\mu \to \nu_e$ 
transition probability.  In section~\ref{sec:ESSnuSB-set-up},  
 the ESS$\nu$SB setup is described in detail. In section~\ref{sec:numerical-procedure} we present
 the details of  our numerical study. In Section~\ref{sec:MH} we briefly explain the (lack of) sensitivity to the neutrino
mass hierarchy and to the octant of $\theta_{23}$ making use of the bievents
plots. In Section~\ref{sec:CPV} we describe the sensitivity to CPV 
and the ability to reconstruct the CP phases. 
Finally, we trace the conclusions in section~\ref{Conclusions}. 

\section{Transition probability in the 4-flavor scheme}
\label{sec:probability}
\subsection{Theoretical framework}

In the enlarged 3+1 framework, the connection among the flavor ($\nu_e,\nu_\mu,\nu_\tau, \nu_s$)
and the mass eigenstates ($\nu_1,\nu_2,\nu_3,\nu_4$) is provided by a $4\times4$ 
unitary matrix
\begin{equation}
\label{eq:U}
U =   \tilde R_{34}  R_{24} \tilde R_{14} R_{23} \tilde R_{13} R_{12}\,, 
\end{equation} 
where $R_{ij}$ ($\tilde R_{ij}$) represents a real (complex) $4\times4$ rotation 
of a mixing angle $\theta_{ij}$ which contains the $2\times2$ submatrix 
\begin{eqnarray}
\label{eq:R_ij_2dim}
     R^{2\times2}_{ij} =
    \begin{pmatrix}
         c_{ij} &  s_{ij}  \\
         - s_{ij}  &  c_{ij}
    \end{pmatrix} ,
\,\,\,\,\,\,\,   
     \tilde R^{2\times2}_{ij} =
    \begin{pmatrix}
         c_{ij} &  \tilde s_{ij}  \\
         - \tilde s_{ij}^*  &  c_{ij}
    \end{pmatrix}
\,,    
\end{eqnarray}
in the  $(i,j)$ sub-block. For brevity we have defined
\begin{eqnarray}
 c_{ij} \equiv \cos \theta_{ij}, \qquad s_{ij} \equiv \sin \theta_{ij}, \qquad  \tilde s_{ij} \equiv s_{ij} e^{-i\delta_{ij}}.
\end{eqnarray}
The parametrization in Eq.~(\ref{eq:U}) is particularly advantageous because: i) The 
3-flavor expression is recovered by setting $\theta_{14} = \theta_{24} = \theta_{34} =0$.
ii) For small values of the mixing angles $\theta_{14}$, $\theta_{24}$, and $\theta_{13}$, 
it is $|U_{e3}|^2 \simeq s^2_{13}$, $|U_{e4}|^2 = s^2_{14}$, 
$|U_{\mu4}|^2  \simeq s^2_{24}$, and $|U_{\tau4}|^2 \simeq s^2_{34}$, 
implying a clear physical meaning of the three mixing angles. 
iii) Positioning the matrix $\tilde R_{34}$ in the 
leftmost location ensures that the $\nu_{\mu} \to \nu_{e}$ conversion probability
in vacuum is independent of $\theta_{34}$ and of the related CP phase $\delta_{34}$
(see~\cite{Klop:2014ima}). 

\subsection{Conversion probability}
\label{subsec:vacuum}

For the ESS$\nu$SB baseline (540 km), matter effects are very small.
This allows us to limit the discussion to the case of propagation in vacuum. 
As first shown in~\cite{Klop:2014ima}, the $\nu_{\mu} \to \nu_{e}$ 
the conversion probability is the sum of three contributions
\begin{eqnarray}
\label{eq:Pme_4nu_3_terms}
P^{4\nu}_{\mu e}  \simeq  P^{\rm{ATM}} + P^{\rm {INT}}_{\rm I}+   P^{\rm {INT}}_{\rm II}\,.
\end{eqnarray}
The first term is positive definite and depends on the atmospheric mass-squared splitting. It
 provides the leading contribution to the transition probability. The expression of this term
is given by
\begin{eqnarray}
\label{eq:Pme_atm}
 &\!\! \!\! \!\! \!\! \!\! \!\! \!\!  P^{\rm {ATM}} &\!\! \simeq\,  4 s_{23}^2 s^2_{13}  \sin^2{\Delta}\,,
 \end{eqnarray}
where $\Delta \equiv  \Delta m^2_{31}L/4E$ is the (atmospheric) oscillating factor, 
$L$  and $E$ being the neutrino baseline and energy, respectively. The other two terms
in Eq.~(\ref{eq:Pme_4nu_3_terms}) are induced by the interference of two different frequencies 
and are not positive definite.. The second term in Eq.~(\ref{eq:Pme_4nu_3_terms})
is related to the  interference of the solar and atmospheric frequencies and can be expressed as
\begin{eqnarray}
 \label{eq:Pme_int_1}
 &\!\! \!\! \!\! \!\! \!\! \!\! \!\! \!\! P^{\rm {INT}}_{\rm I} &\!\!  \simeq\,   8 s_{13} s_{12} c_{12} s_{23} c_{23} (\alpha \Delta)\sin \Delta \cos({\Delta + \delta_{13}})\,.
\end{eqnarray}
It should be noticed that at the first (second) oscillation maximum one has $\Delta \sim \pi/2$ ($\Delta \sim 3\pi/2$).
For this reason, in ESS$\nu$SB, which works at the second oscillation maximum, one expects an
enhanced sensitivity to the CP phase $\delta_{13}$. Indeed, in spite of the lower statistics,
we will see how ESS$\nu$SB can attain a sensitivity similar to that obtained in the higher statistics experiment T2HK,
which works at the first oscillation maximum. The third term in Eq.~(\ref{eq:Pme_4nu_3_terms})
appears as a new genuine 4-flavor effect, and is connected to the interference of 
sterile and atmospheric frequencies. It can be written in the form~\cite{Klop:2014ima} 
\begin{eqnarray}
 \label{eq:Pme_int_2}
 &\!\! \!\! \!\! \!\! \!\! \!\! \!\! \!\! P^{\rm {INT}}_{\rm II} &\!\!  \simeq\,   4 s_{14} s_{24} s_{13} s_{23} \sin\Delta \sin (\Delta + \delta_{13} - \delta_{14})\,.
\end{eqnarray}
From Eqs.~(\ref{eq:Pme_atm})-(\ref{eq:Pme_int_2}), we can observe that the transition probability 
depends upon three small mixing angles: the standard angle $\theta_{13}$ and two new angles $\theta_{14}$ and $\theta_{24}$.
We notice that the estimates of such three mixing angles (calculated in the 3-flavor 
framework~\cite{deSalas:2017kay,Capozzi:2018ubv,Esteban:2018azc} for $\theta_{13}$,
and in the 4-flavor scheme~\cite{Capozzi:2016vac,Gariazzo:2017fdh,Dentler:2018sju,Diaz:2019fwt} 
for $\theta_{14}$ and $\theta_{24}$) are similar and one has 
$s_{13} \sim s_{14} \sim s_{24} \sim 0.15$ (see table~\ref{tab:benchmark-parameters}). 
Therefore, one can consider these three angles as small parameters  having 
the same order $\epsilon$. We note also that the ratio of the solar over the 
atmospheric mass-squared splittings, 
$\alpha \equiv \Delta m^2_{21}/ \Delta m^2_{31} \simeq \pm \, 0.03$
can be treated as of order $\epsilon^2$. 
From Eqs.~(\ref{eq:Pme_atm})-(\ref{eq:Pme_int_2}), we deduce that the 
first (leading) contribution is of the second order, while the two interference terms 
are of the third one. However, differently from the standard 
interference term in Eq.~(\ref{eq:Pme_int_1}), the new sterile induced
interference term in Eq.~(\ref{eq:Pme_int_2}) is not proportional 
to $\Delta$, so it is not enhanced at the second oscillation maximum.
Because of this feature, as it will be confirmed by our numerical simulations,
the performance of ESS$\nu$SB in the 3+1 scheme is not as good as that
of those experiments which work at the first oscillation maximum, such as T2HK and DUNE.

\section{Experimental Specifications}
\label{sec:ESSnuSB-set-up}
In this section, we briefly discuss the specifications of the experimental setup ESS$\nu$SB. ESS$\nu$SB is a proposed superbeam on-axis experiment where a very high intense proton beam of energy 2 GeV with an average beam power of 5 MW will be delivered by the European Spallation Source (ESS) linac facility running at 14 Hz. The number of protons on target (POT) per year (208 days) will be 2.7$\times 10^{23}$~\cite{Baussan:2013zcy,Dracos:2016wso,Dracos:2018jsn,Dracos:2018syh}. It is worth to mention here that the future linac upgrade can push the proton energy up to 3.6 GeV. This highly ambitious and exciting facility is expected to start taking neutrino data around 2030. We have obtained the fluxes from \cite{enrique} and these on-axis (anti)neutrino fluxes arising from the 2 GeV protons on target peaks around 0.25 GeV. In this case a 500 kt fiducial mass Water Cherenkov detector similar to the properties of the MEMPHYS detector \cite{Agostino:2012fd,luca} has been proposed to explore the neutrino properties in this low energy regimes. It has been shown in \cite{Baussan:2013zcy} that if the detector is placed in any of the existing mines located in between 300-600 km from the ESS site at Lund, it will make possible to achieve 5$\sigma$ confidence level discovery of leptonic CP-violation up to the 50\% coverage of the whole range of CP phases. A detailed study on the CP-violation discovery capability of this facility with different baseline and different combinations of neutrino and antineutrino run time has also been explored in \cite{Agarwalla:2014tpa}. In this work, we consider a baseline of 540 km from Lund to Garpenberg mine located in Sweden and also we have matched the event numbers of Table~3 and all other results given in~\cite{Baussan:2013zcy}. At this baseline, it fully covers the second oscillation maximum and it provides the opportunity to explore the CP-asymmetry which (in the 3-flavor scheme) is three times larger than the CP-asymmetry at the first oscillation maximum. Although the main drawbacks for going to the second oscillation maximum come from the significant decrease of statistics and cross-sections compared to the first oscillation maximum, the high intense beam of this excellent facility takes care of those difficulties and make the statistics competitive to provide exciting results. All our simulations presented here for this setup have been done assuming 2 yrs of $\nu$ and 8 yrs of $\bar{\nu}$ running with a most optimistic consideration of uncorrelated 5\% signal normalization and 10\% background normalization error for both neutrino and antineutrino appearance and disappearance channels respectively. For more details of the accelerator facility, beamline design, and detector facility of this setup please see~\cite{Baussan:2013zcy}.

\section{Details of the Numerical Analysis}
\label{sec:numerical-procedure}

\begin{table}[t]
\begin{center}
{
\newcommand{\mc}[3]{\multicolumn{#1}{#2}{#3}}
\newcommand{\mr}[3]{\multirow{#1}{#2}{#3}}
\begin{tabular}{|c|c|c|}
\hline\hline
\mr{2}{*}{\bf Parameter} & \mr{2}{*}{\bf True Value} & \mr{2}{*}{\bf Marginalization Range} \\
  & &  \\
\hline\hline
\mr{2}{*}{$\sin^2{\theta_{12}}$} & \mr{2}{*}{0.304} & \mr{2}{*}{Not marginalized} \\
  & &  \\
\hline
\mr{2}{*}{$\sin^22\theta_{13}$} & \mr{2}{*}{$0.085$} & \mr{2}{*}{Not marginalized} \\ 
  & &  \\
\hline
\mr{2}{*}{$\sin^2{\theta_{23}}$} & \mr{2}{*}{0.50} & \mr{2}{*}{[0.34, 0.68]} \\
  & &  \\
\hline
\mr{2}{*}{$\sin^2{\theta_{14}}$} & \mr{2}{*}{0.025} & \mr{2}{*}{Not marginalized} \\
  & &  \\  
\hline
\mr{2}{*}{$\sin^2{\theta_{24}}$} & \mr{2}{*}{0.025} & \mr{2}{*}{Not marginalized} \\
  & &  \\ 
\hline
\mr{2}{*}{$\sin^2{\theta_{34}}$} & \mr{2}{*}{0.0} & \mr{2}{*}{Not marginalized} \\
  & &  \\  
\hline
\mr{2}{*}{$\delta_{13}/^{\circ}$} & \mr{2}{*}{[- 180, 180]} & \mr{2}{*}{[- 180, 180]} \\
  & &  \\
\hline
\mr{2}{*}{$\delta_{14}/^{\circ}$} & \mr{2}{*}{[- 180, 180]} & \mr{2}{*}{[- 180, 180]} \\
  & &  \\
\hline
\mr{2}{*}{$\delta_{34}/^{\circ}$} & \mr{2}{*}{0} & \mr{2}{*}{Not marginalized} \\
  & &  \\  
\hline
\mr{2}{*}{$\frac{\Delta{m^2_{21}}}{10^{-5} \, \rm{eV}^2}$} & \mr{2}{*}{7.50} & \mr{2}{*}{Not marginalized} \\
  & &  \\
\hline
\mr{2}{*}{$\frac{\Delta{m^2_{31}}}{10^{-3} \, \rm{eV}^2}$ (NH)} & \mr{2}{*}{2.475} &\mr{2}{*}{Not marginalized} \\
  & & \\
\hline
\mr{2}{*}{$\frac{\Delta{m^2_{31}}}{10^{-3} \, \rm{eV}^2}$ (IH)} & \mr{2}{*}{- 2.4} &\mr{2}{*}{Not marginalized} \\
  & & \\
\hline
\mr{2}{*}{$\frac{\Delta{m^2_{41}}}{\rm{eV}^2}$} & \mr{2}{*}{1.0} & \mr{2}{*}{Not marginalized} \\
  & &  \\
\hline\hline
\end{tabular}
}
\caption{Oscillation parameters along with their true values and marginalization status shown in this table. The second column
represents the values of the parameters used to generate the true 
data set. The third column displays the parameters which are kept fixed in the fit and the parameters which have been marginalized in the fit within their allowed ranges.}
\label{tab:benchmark-parameters}
\end{center}
\end{table}

This section details the numerical analysis adopted to produce the sensitivity results presented in the following sections. To compute the sensitivity measurements along with the bi-events plots we have used the GLoBES software~\cite{Huber:2004ka,Huber:2007ji} and its new tool~\cite{Kopp:NSI} which can include the sterile neutrinos. In this paper, we have adopted the same strategy for the simulation described in section 4 of ref.~\cite{Agarwalla:2016mrc}.  The true values of the oscillation parameters together  with their marginalization ranges considered in our simulations are presented in Table~\ref{tab:benchmark-parameters}. Our  benchmark choices for the three-flavor neutrino oscillation parameters closely resemble those obtained in the latest global fits~\cite{Capozzi:2018ubv,deSalas:2017kay,Esteban:2018azc}, although we have made the true choice of the atmospheric mixing angle to be maximal ($45^{\circ}$)%
\footnote{Recent 3$\nu$ global fits~\cite{Capozzi:2018ubv,deSalas:2017kay,Esteban:2018azc} sligthy prefer non-maximal $\theta_{23}$ with two nearly degenerate solutions: one is $<$ $45^{\circ}$, in the lower octant (LO), and the other is $>$ $45^{\circ}$, in the higher octant (HO). However, maximal mixing is still allowed at 2$\sigma$ confidence level.}, and in the fit, it has been marginalized over its allowed range as mentioned in the third column of Table~\ref{tab:benchmark-parameters}. 
Concerning the active-sterile mixing angles we have taken the benchmark values very close to those obtained in the global fit analyses~\cite{Capozzi:2016vac,Gariazzo:2017fdh,Dentler:2018sju,Diaz:2019fwt} performed within the 3+1 scheme%
\footnote{We stress that assuming smaller values for $\theta_{14}$ and $\theta_{24}$ the impact of active-sterile oscillations would decrease. As a consequence
the sensitivity to CPV induced by $\delta_{14}$ would be reduced. On the other hand, the deterioration of the sensitivity to the CPV induced by the standard CP phase $\delta_{13}$ would be less.}.

 In all our simulations, we have assumed normal hierarchy%
\footnote{We have checked that the results with the true choice of inverted hierarchy are similar to the results presented in this work.} 
(NH) as the true choice and we have kept it fixed also in the fit. In fact, we are assuming that the correct hierarchy will be already known by the time ESS$\nu$SB will start to
take data. The two mixing angles $\theta_{12}$ and $\theta_{13}$ have been kept fixed in the data as well as in the fit taking into account the stringent constraints provided by the solar and the reactor data. We have also kept the two mass-squared differences $\Delta m_{21}^2$ and $\Delta m_{31}^2$ fixed at their true choices and they have not been marginalized in the fit. $\delta_{13}$ (true) has been taken from its allowed range of $[-\pi, \pi]$, while in the fit we have marginalized over its 
full range depending on the analysis requirement. In our simulations, we consider the constant line-averaged Earth matter density of 2.8 g/cm$^{3}$ following the Preliminary Reference Earth Model (PREM)~\cite{PREM:1981}. The new mass-squared splitting $\Delta m^2_{41}$ arising in the 3+1 scheme is taken as 1 eV$^2$ following the present preference of the short-baseline data%
\footnote{We stress that our sensitivity results would remain unaltered provided $\Delta m^2_{41} \gtrsim 0.1\,$eV$^2$.}.
This large value of $\Delta m^2_{41}$ induces fast oscillations which get averaged out due to the finite energy resolution of the detector. As a result the sign of $\Delta m^2_{41}$ is irrelevant in this setup. Now, the new mixing angles $\theta_{14}$ and $\theta_{24}$ emerging out of the 3+1 framework, have been taken fixed at their true values in the data as well as in the fit%
\footnote{We point out that our choice to fix the fit values of $\theta_{14}$ and $\theta_{24}$ is well justified if one assumes (as we do) that one has precise information 
on these two parameters coming form SBL experiments. In such a case, the marginalization of $\theta_{14}$ and $\theta_{24}$  in the fit would provide
minor modifications to our results.}. 

The true value of the new CP phase $\delta_{14}$  is taken in its allowed range [$-\pi$, $\pi$] and its test  value has been marginalized over the allowed range if required. The mixing angle $\theta_{34}$ has been considered to be zero both in the data and in theory. This choice makes the presence of its associated phase $\delta_{34}$ irrelevant in the simulation%
\footnote{According to our parametrization followed in Eq.~(\ref{eq:U}), the $\nu_{\mu} \to \nu_e$ oscillation probability in vacuum is independent of $\theta_{34}$ (and $\delta_{34}$). However it has a higher order ($\epsilon^4$) impact in presence of matter effect, which in case of ESS$\nu$SB baseline is very small. Hence $\theta_{34}$ (and $\delta_{34}$) can safely be ignored in the simulation. A detailed discussion including some analytical understanding regarding this issue is given in the appendix of 
ref.~\cite{Klop:2014ima}.}.

In our analysis, we do not consider any near detector of ESS$\nu$SB which may help to reduce the systematic uncertainties and might give some information on 
the two mixing angles $\theta_{14}$ and $\theta_{24}$. However, it would give no information regarding the active and sterile CP phases which is our main issue
of interest in the present work. It is worth to underline here that in all our simulations we have performed a spectral analysis making use of the binned events spectra. 
In the statistical analysis we not only marginalize over the oscillation parameters but also over the nuisance parameters adopting the well-known ``pull'' method~\cite{Huber:2002mx,Fogli:2002pt} to calculate the Poissonian $\Delta\chi^{2}$. We display our results in terms of the squared-root of $\Delta\chi^{2}$ which represents $n\sigma$ ($n\equiv\sqrt{\Delta\chi^{2}}$) confidence level statistical significance for one degrees of freedom (d.o.f). 

\section{Mass ordering and $\theta_{23}$ octant sensitivity in the 4-flavor scheme}
\label{sec:MH}

 \begin{figure}[t!]
\centerline{
\includegraphics[height=8.cm,width=8.cm]{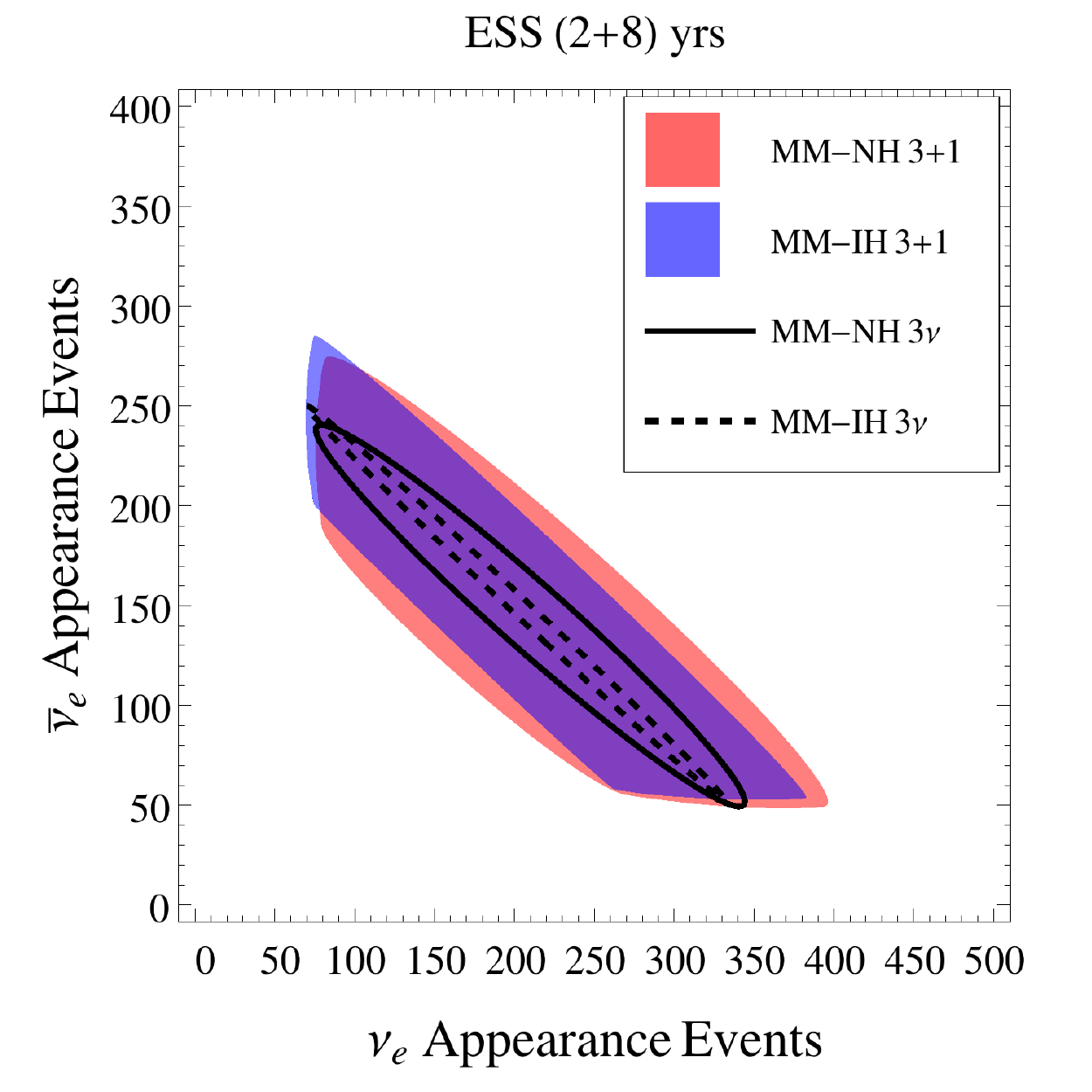}
 \includegraphics[height=8.cm,width=8.cm]{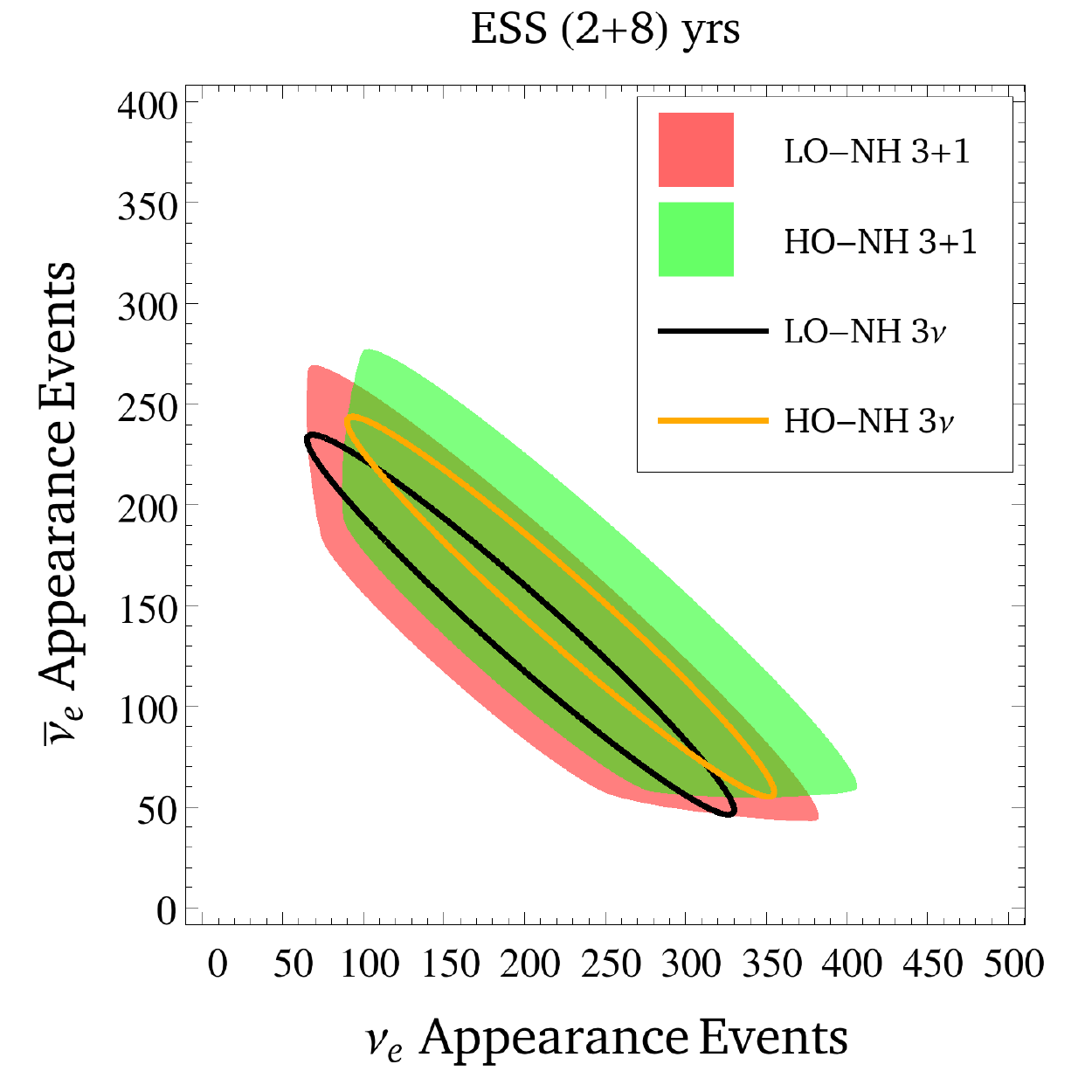}
  }
 \caption{The left panel reports the bievents plot for the 3-flavor framework (black curves) and 3+1 scheme (colored blobs)
 for the case of NH and maximal $\theta_{23}$. The right panel represents, for the NH case, the
 bievents plots corresponding to two values of $\theta_{23}$ in the two octants. We have taken $\sin^2\theta_{23} =0.42\, (0.58)$.
See the text for details. }
 \label{CPV_bievents}
 \end{figure}

It is well known that in the 3-flavor scheme ESS$\nu$SB has scarce sensitivity 
to both these two properties. Concerning the 
MH hierarchy, the lack of sensitivity is due to the fact that matter effects are very small
in ESS$\nu$SB. The low sensitivity to the octant of $\theta_{23}$ is imputable
to the fact that ESS$\nu$SB works at the second oscillation maximum, which
is more narrow than the first one. These features, together with the lower statistics, render ESS$\nu$SB much
less sensitive than other experiments (T2HK for example) to the octant of $\theta_{23}$.

Here we confirm similar findings also in the 4-flavor scheme.
This conclusion can be easily understood through a discussion at the level of the 
neutrino and antineutrino appearance events. The left panel of Fig.~\ref{CPV_bievents} reports the bievent 
plots, where the $x$-axis represents the number of $\nu_e$ events and the $y$-axis represents the $\bar \nu_e$ events.
The two ellipses represent the 3-flavor model and can be obtained by varying 
the CP phase $\delta_{13}$ in the range $[-\pi,\pi]$. The solid (dashed) ellipse
corresponds to the NH (IH). The centroids of the two ellipses basically coincide,
hence it is clear that the setup cannot discriminate the MH. 
This is qualitatively different with respect to what occurs in other LBL experiments 
(T2HK~\cite{Agarwalla:2018t2hk} and especially DUNE~\cite{Agarwalla:2016xxa}),
where the two ellipses get separated due to the presence of the matter effects.
In the 3+1 scheme, two CP phases are present and their variation in the
range $[-\pi,\pi]$ gives even more freedom. The bievent plots obtained 
varying both the CP phases $\delta_{13}$ and $\delta_{14}$ are 
represented by colored elongated blobs in the left panel Fig.~\ref{CPV_bievents}.
The two blobs corresponding to the two hierarchies are completely overlapped.
This implies that, similarly to the 3-flavor scheme, one does not expect any sensitivity
to the MH in the 3+1 scheme as well. 

The right panel of Fig.~\ref{CPV_bievents} reports the ellipses (blobs) obtained in the 3-flavor (4-flavor) cases 
for two values of $\theta_{23}$ chosen in the two opposite octants.
We have taken $\sin^2\theta_{23} =0.42\, (0.58)$ as benchmark values.
We observe that in both schemes there is a partial overlapping between
the regions representing the two octants. The degree of overlapping increases
when going from 3-flavor to the 3+1 scheme. Therefore, we expect a poor
sensitivity to the octant of $\theta_{23}$ both in $3\nu$ and $4\nu$ schemes.
Differently from T2HK, the spectral information is not of great help due to the low statistics.
This is confirmed by the numerical simulations (not shown) performed by 
including the full energy spectrum in the fit.

\section{CP-violation searches in the 4-flavor framework}
\label{sec:CPV}
   
In this section, we analyze the capability of ESS$\nu$SB of pinning down the extended
CPV sector entailed by the 3+1 scheme. First we assess the sensitivity
to the CPV induced by the CP phase $\delta_{13}$ and $\delta_{14}$. Second 
we discuss the capability of reconstructing the true values of the two phases $\delta_{13}$ and $\delta_{14}$.

 \begin{figure}[t!]
\centerline{
\includegraphics[height=8.cm,width=8.cm]{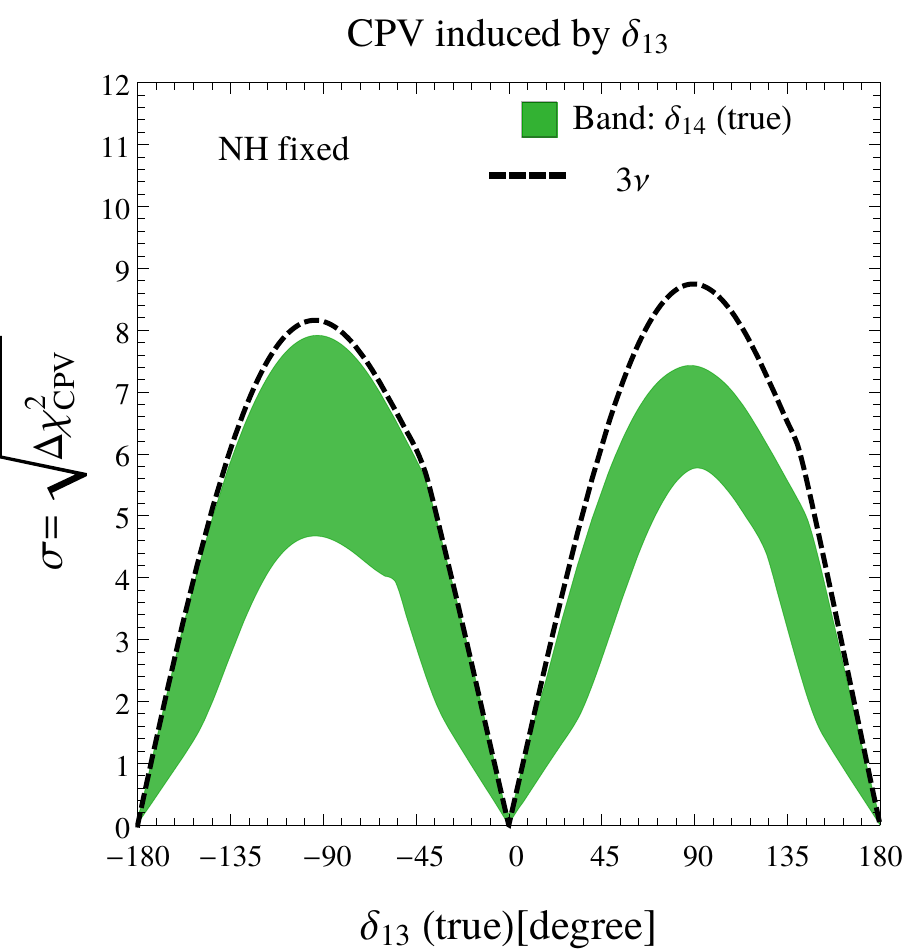}
 \includegraphics[height=8.cm,width=8.cm]{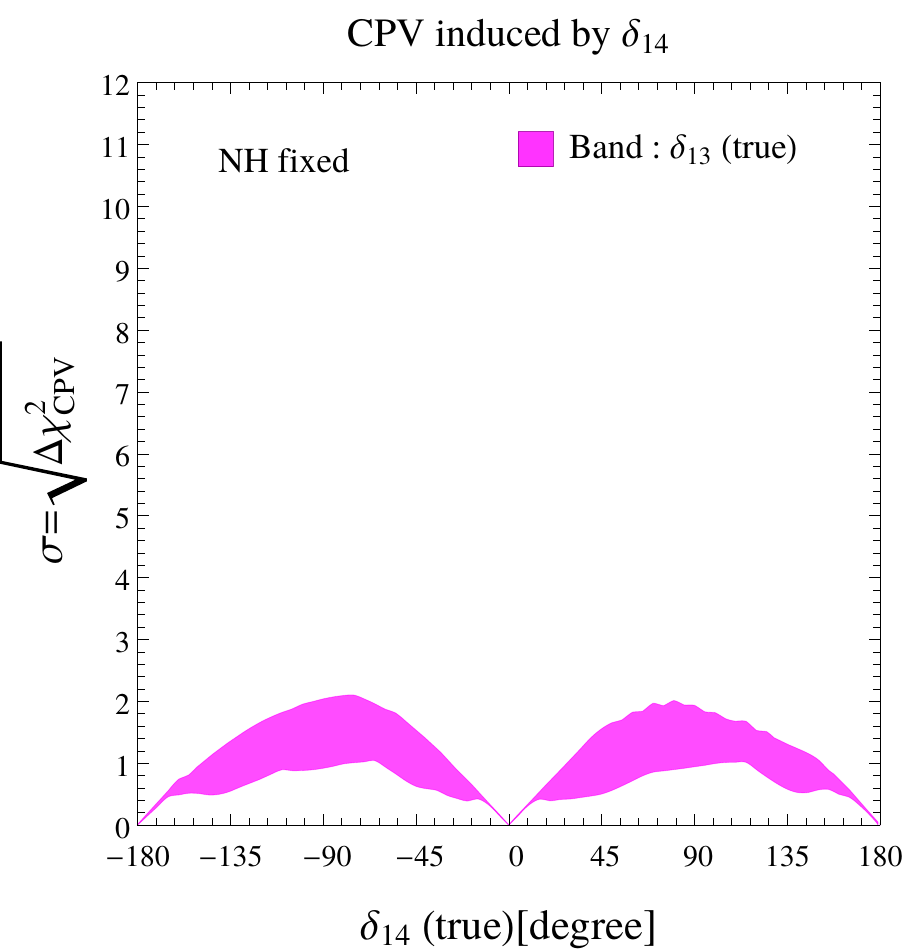}
  }
\caption{ESS$\nu$SB discovery potential of  $\delta_{13} \ne (0,\pi)$ (left panel)
and  $\delta_{14} \ne (0,\pi)$ (right panel). In both panels
  the MH is fixed to be the NH (both true and test value). 
 The black dashed curve corresponds to the 3-flavor case while
 the colored band correspond to the 3+1 scheme.  In this last case, we have fixed 
 the true and test values of $\theta_{14}$ = $\theta_{24}$ = $9^0$ and varied the unknown 
value of the true $\delta_{14}$ in its entire range of $[-\pi,\pi]$ while marginalizing over test 
$\delta_{14}$ in the same range.}
 \label{CPV_delta_13}
 \end{figure}

\subsection{Sensitivity to CP-violation}

The sensitivity of CPV produced by a fixed (true) value of a CP phase $\delta_{ij}^{\mathrm{true}}$ 
can be defined as the statistical significance at which one can reject the test hypothesis of no CPV, 
i.e. the two (test) cases $\delta_{ij}^{\mathrm{test}}=0,\pi$.
 In the left panel of Fig.~\ref{CPV_delta_13}, we report the discovery potential of CPV induced by $\delta_{13}$. 
We have assumed that the hierarchy is known a priori and is NH. 
The dashed black curve correspond to the 3-flavor scheme
while the green band to the 3+1 scheme. In the 3+1 scenario, we fix the test and true 
values of $\theta_{14} =9^0$ and $\theta_{24}=9^0$. The green
band is attained by varying the unknown true value of $\delta_{14}$
in the range of $[-\pi,\pi]$ and marginalizing over its test values. 
We observe that in the 3+1 scheme there is a deterioration of the sensitivity.
Adopting $\delta_{13} = -90^0$ as a benchmark value in the 3-flavor (4-flavor)
scheme one has 8.2$\sigma$ (4.5$\sigma$) sensitivity. We find very similar result for the case of IH
(not shown). The right panel of Fig.~\ref{CPV_delta_13} displays the discovery potential of CPV induced 
by $\delta_{14}$ for the NH case. The magenta band is obtained by varying the true values of
the CP phase $\delta_{13}$ in the range $[-\pi,\pi]$  while marginalizing over their
test values in the same range in the fit. We observe that ESS$\nu$SB has a limited sensitivity
to the CP phase $\delta_{14}$, which is always below the 2$\sigma$ level. 

We think that it is useful to make a comparison between the results obtained here for 
ESS$\nu$SB with those found for T2HK in our work~\cite{Agarwalla:2018t2hk}. 
We notice that in the 3-flavor scheme both experiments have a similar sensitivity 
to CPV induced by $\delta_{13}$, having both a maximal sensitivity of about 8$\sigma$ for
the values $\delta_{13}\simeq \pm 90^0$. This is possible because, despite of the lower
statical power, ESS$\nu$SB benefits of the amplification factor proportional to $\Delta$,
which is three times bigger at the second oscillation maximum with respect to the first one. 
In contrast, in the presence of a sterile
neutrino, the performance is much worse in ESS$\nu$SB. In fact, one can notice 
the two following features:  i) The deterioration of the sensitivity
to the CPV driven by $\delta_{13}$ when going from the 3-flavor to the 3+1 scheme is much more
pronounced in ESS$\nu$SB than in T2HK. Taking the values $\delta_{13} = \pm 90^0$ as a benchmark
(where the maximal sensitivity is attained) in~\cite{Agarwalla:2018t2hk}, we found for T2HK only a weak reduction
of the sensitivity from 8$\sigma$ to 7$\sigma$ (see Fig.~4 in~\cite{Agarwalla:2018t2hk}). In ESS$\nu$SB, we now find a
severe reduction from  8$\sigma$ to 4.5$\sigma$ (see left panel of Fig.~\ref{CPV_delta_13});
ii) The sensitivity to the CPV induced by
the CP phase $\delta_{14}$ is considerably lower in ESS$\nu$SB than in T2HK ($2\sigma$ vs  $5\sigma$
for $\delta_{14} = \pm 90^0$). The explanation of such
a different performance in the 3+1 scheme of the two experiments
can be traced to the fact that T2HK (ESS$\nu$SB) works around the first 
(second) oscillation maximum.  As already noticed in subection~\ref{subsec:vacuum},
the new interference term (which depends on $\delta_{14}$), at the second oscillation
maximum is not amplified by the factor $\Delta$ as it happens for the standard interference
 term (which depends on $\delta_{13}$). In addition, as remarked in~\cite{Agarwalla:2018t2hk}
 in T2HK the spectral information
plays a crucial role in guaranteeing a good performance in the 3+1 scheme. Indeed,
in~\cite{Agarwalla:2018t2hk}, we explicitly showed that even if there is a complete degeneracy at the level of the event counting,
the energy spectrum provides additional information which breaks such a degeneracy
and boosts the sensitivity. In ESS$\nu$SB, the role of the spectral information is substantially 
reduced because of the energy range at the second oscillation maximum is very narrow and 
the low statistics does not allow to exploit the information contained in the spectrum. 
Hence, we conclude that ESS$\nu$SB is not particularly
suited for the CPV related searches in the presence of sterile neutrinos. 
 
 \begin{figure}[t!]
\centering
 \includegraphics[height=4.9cm,width=4.9cm]{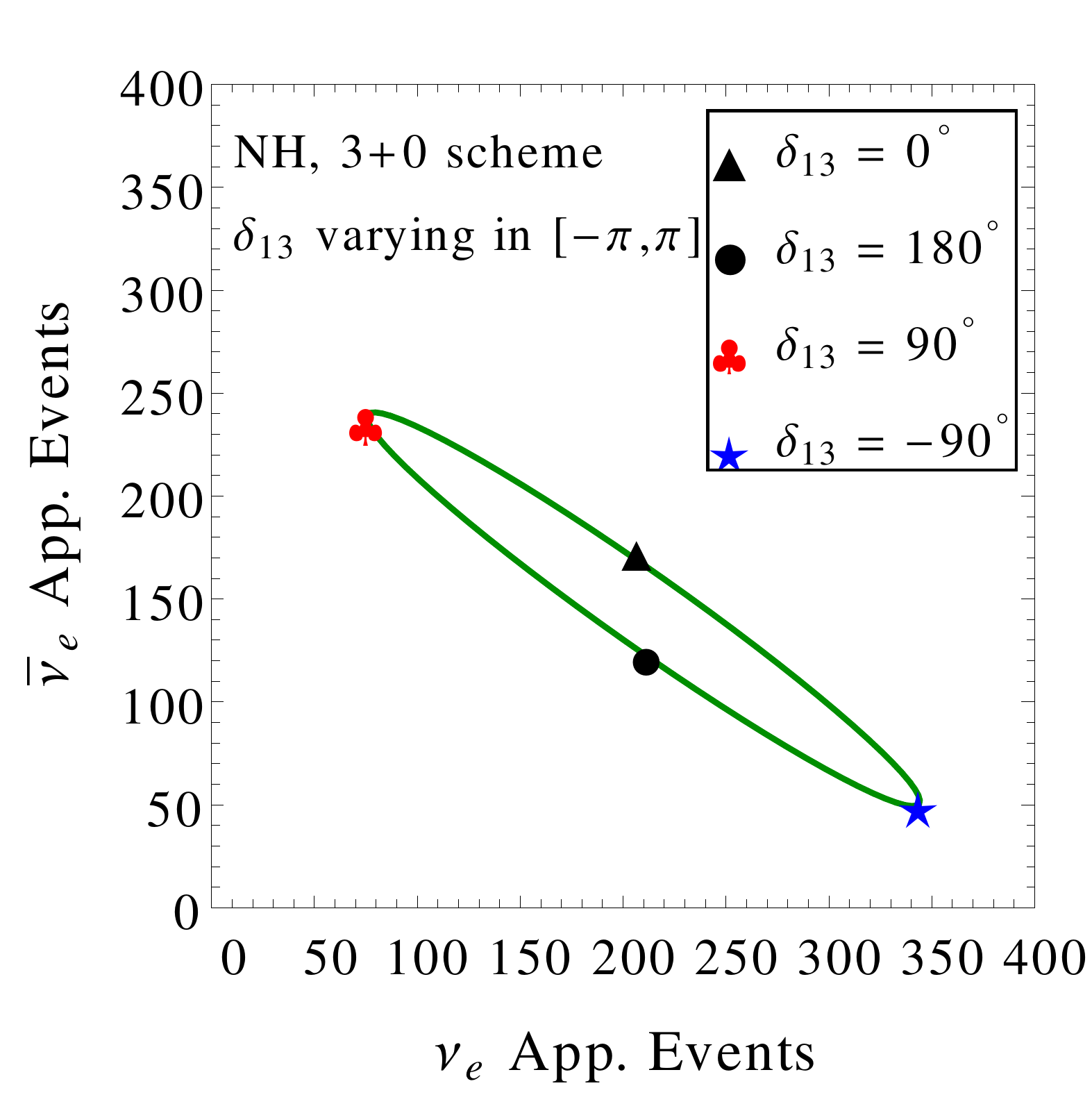}
 \includegraphics[height=4.9cm,width=4.9cm]{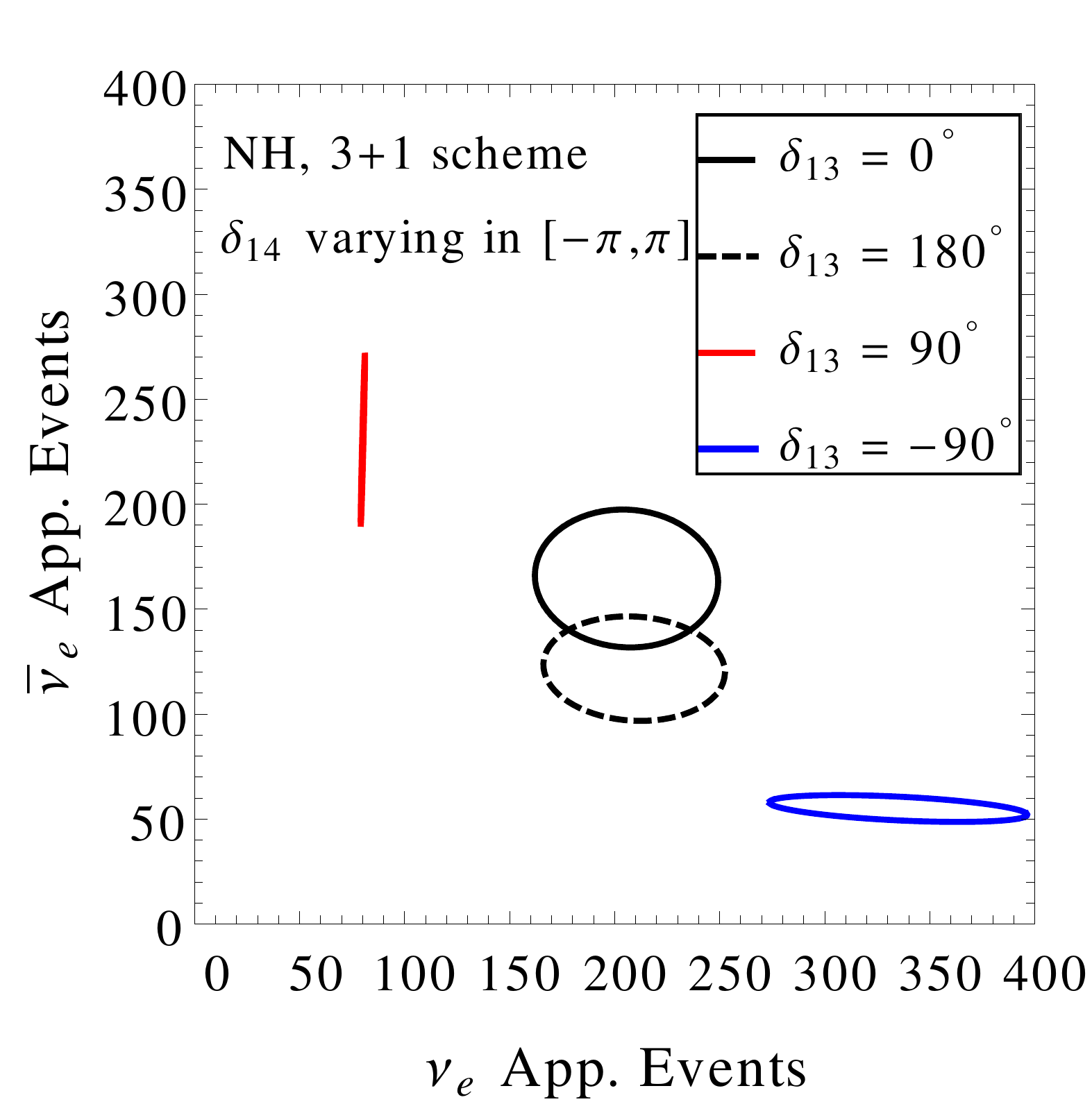}
 \includegraphics[height=4.9cm,width=4.9cm]{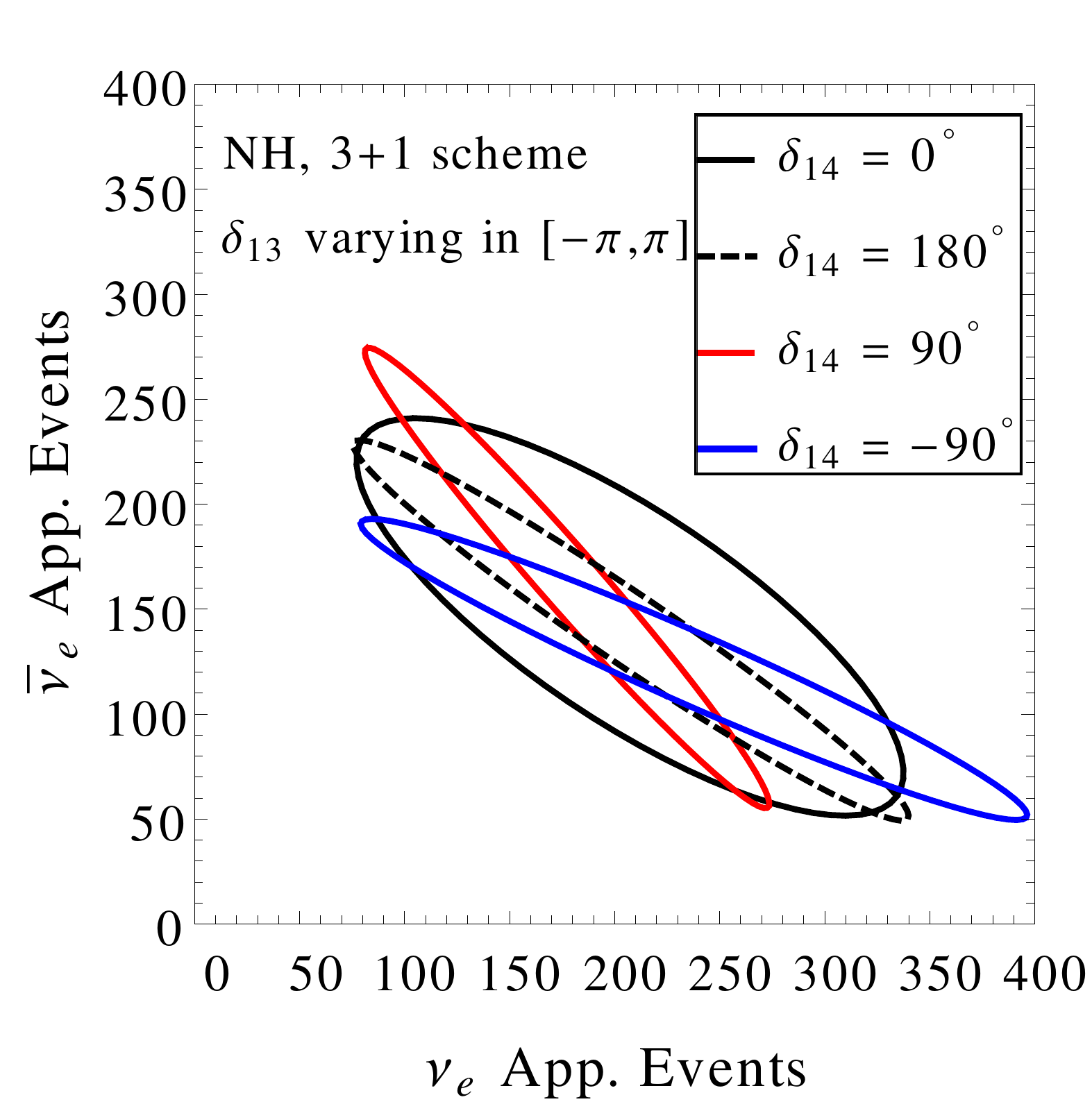}
 \caption{The left panel refers to the standard 3-flavor framework. In this case the model 
 lies on the (green) ellipse, which is obtained by varying $\delta_{13}$ in the range $[-\pi,\pi]$. 
 The two black marks represent the cases of no CPV ($\delta_{13} = 0,\pi$) while the two
  colored ones correspond to the cases of maximal CPV  ($\delta_{13} = -\pi/2,\pi/2$).   
  The non-zero distance between the black marks and the colored ones implies that events 
  counting can detect the CPV induced by the phase $\delta_{13}$. The second panel refers to
  the 3+1 scheme. In this case a fixed value of $\delta_{13}$ is represented by an ellipse,
  where $\delta_{14}$ varies in the range $[-\pi,\pi]$.
  The non-zero distance between the black ellipses and the two colored ones implies
  that events counting is sensitive to CPV induced by the phase $\delta_{13}$ also in the 3+1 case. 
  However, the distances are reduced with respect to the 3-flavor case. Therefore, the sensitivity      
  decreases. The right panel refers to the 3+1 scheme and illustrates the sensitivity to the CPV induced by $\delta_{14}$. In this case  we plot four ellipses corresponding to the four values of $\delta_{14}$ (while $\delta_{13}$ is varying in the range $[-\pi,\pi]$).
   Each of the two ellipses (blue and red) 
  corresponding to maximal CPV induced by $\delta_{14}$ intercepts  
  the two ellipses (solid and dashed black) corresponding to no CPV induced by $\delta_{14}$. 
   In the crossing points the events counting is completely insensitive to CPV induced by the new CP phase $\delta_{14}$. }
     \label{fig:bievent_3panel}
 \end{figure}

The situation can be further clarified by inspecting the 3-panel bievent plot displayed in 
Fig.~\ref{fig:bievent_3panel}. The left panel refers to the standard 3-flavor framework. In this case the model lies on the (green) ellipse, which is obtained by varying $\delta_{13}$ in the range $[-\pi,\pi]$. 
 The two black marks represent the cases of no CPV ($\delta_{13} = 0,\pi$) while the two
  colored ones correspond to the cases of maximal CPV  ($\delta_{13} = -\pi/2,\pi/2$).   
  The non-zero distance between the black marks and the colored ones implies that events 
  counting can detect the CPV induced by the phase $\delta_{13}$. The second panel refers to
  the 3+1 scheme. In this case a fixed value of $\delta_{13}$ is represented by an ellipse,
  where $\delta_{14}$ varies in the range $[-\pi,\pi]$.
  The non-zero distance between the black ellipses and the two colored ones implies
  that events counting is sensitive to CPV induced by the phase $\delta_{13}$ also in the 3+1 case. 
  However, the distances are reduced with respect to the 3-flavor case. Therefore, the sensitivity      
  decreases as found in the numerical simulation as shown in the left panel of Fig.~\ref{CPV_delta_13}.  
  The right panel of Fig.~\ref{fig:bievent_3panel} refers to the 3+1 scheme and illustrates the sensitivity to the CPV induced by $\delta_{14}$. In this case  we plot four ellipses corresponding to the four values of $\delta_{14}$ (while $\delta_{13}$ is varying in the range $[-\pi,\pi]$).
   Each of the two ellipses (blue and red) 
  corresponding to maximal CPV induced by $\delta_{14}$ intercepts  
  the two ellipses (solid and dashed black) corresponding to no CPV induced by $\delta_{14}$. 
   In the crossing points the events counting is completely insensitive to CPV induced by the new CP phase $\delta_{14}$.  Therefore there are always (unlucky) combinations of the CP phases for which the event counting cannot determine if there is CPV induced by $\delta_{14}$. Notwithstanding
in the right panel of Fig.~\ref{CPV_delta_13}, we observe that there is $\sim 2\sigma$ sensitivity
for $\delta_{14} = \pm 90^0$. We have checked that such a residual sensitivity comes from the
spectral shape information. As already remarked above, this information in ESS$\nu$SB is
much weaker compared to T2HK, and as a consequence the sensitivity remains quite low.
  
\begin{figure}[t!]
\hspace{0.5cm}
\centerline{
\includegraphics[height=14.0cm,width=14.0cm]{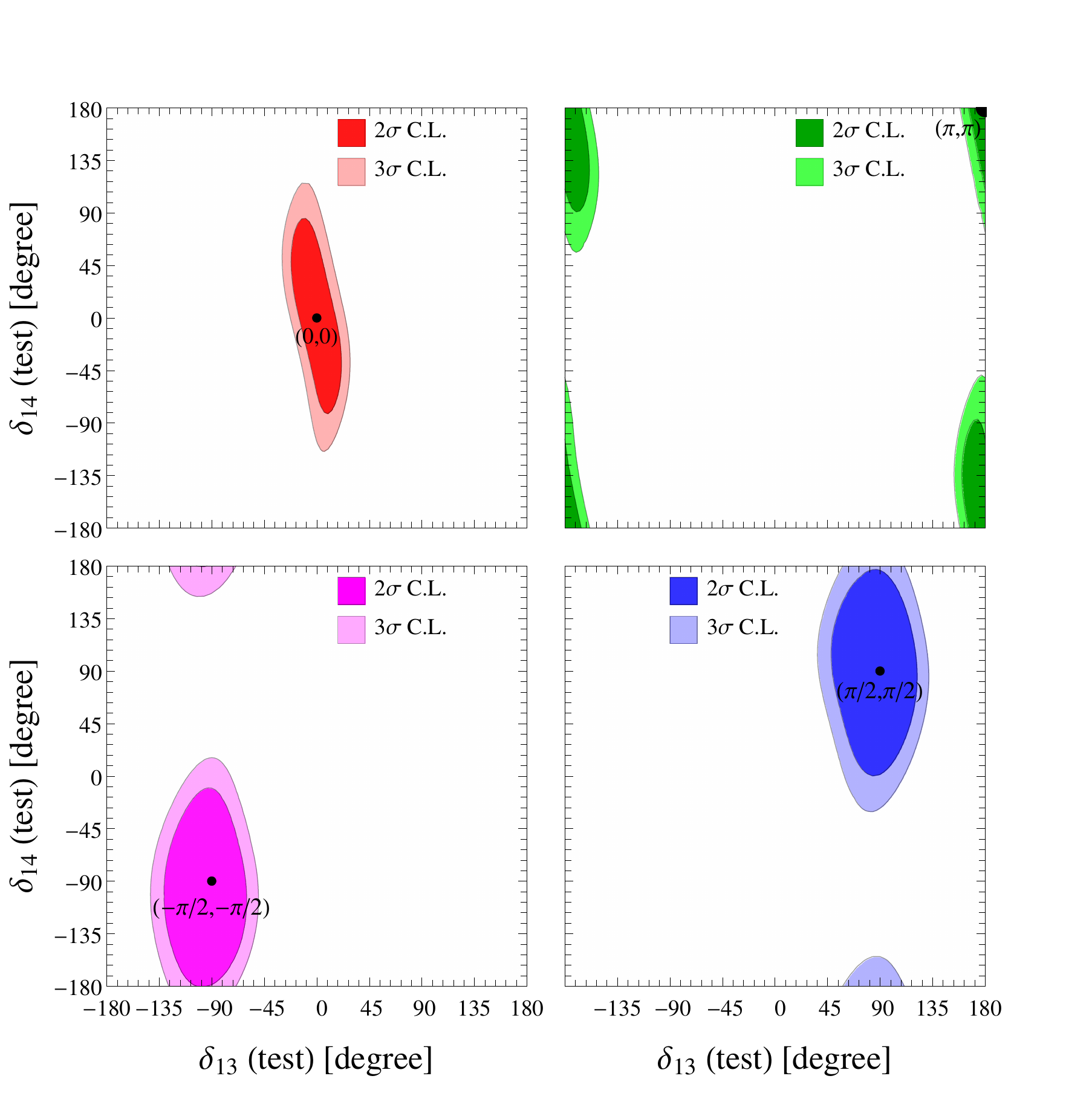}}
\vspace{-0.5cm}
\caption{Reconstructed regions for the two CP phases $\delta_{13}$ and $\delta_{14}$ 
for the four benchmark pairs of their true values indicated in each panel. We have fixed
the NH as the true and test hierarchy. The contours refer to 2$\sigma$ and 3$\sigma$ 
confidence levels (1 d.o.f.).}
\label{CPV_rec_1}
\end{figure}
 
\begin{figure}[t!]
\hspace{-0.2cm}
\centerline{
\includegraphics[height=15.0cm,width=15.0cm]{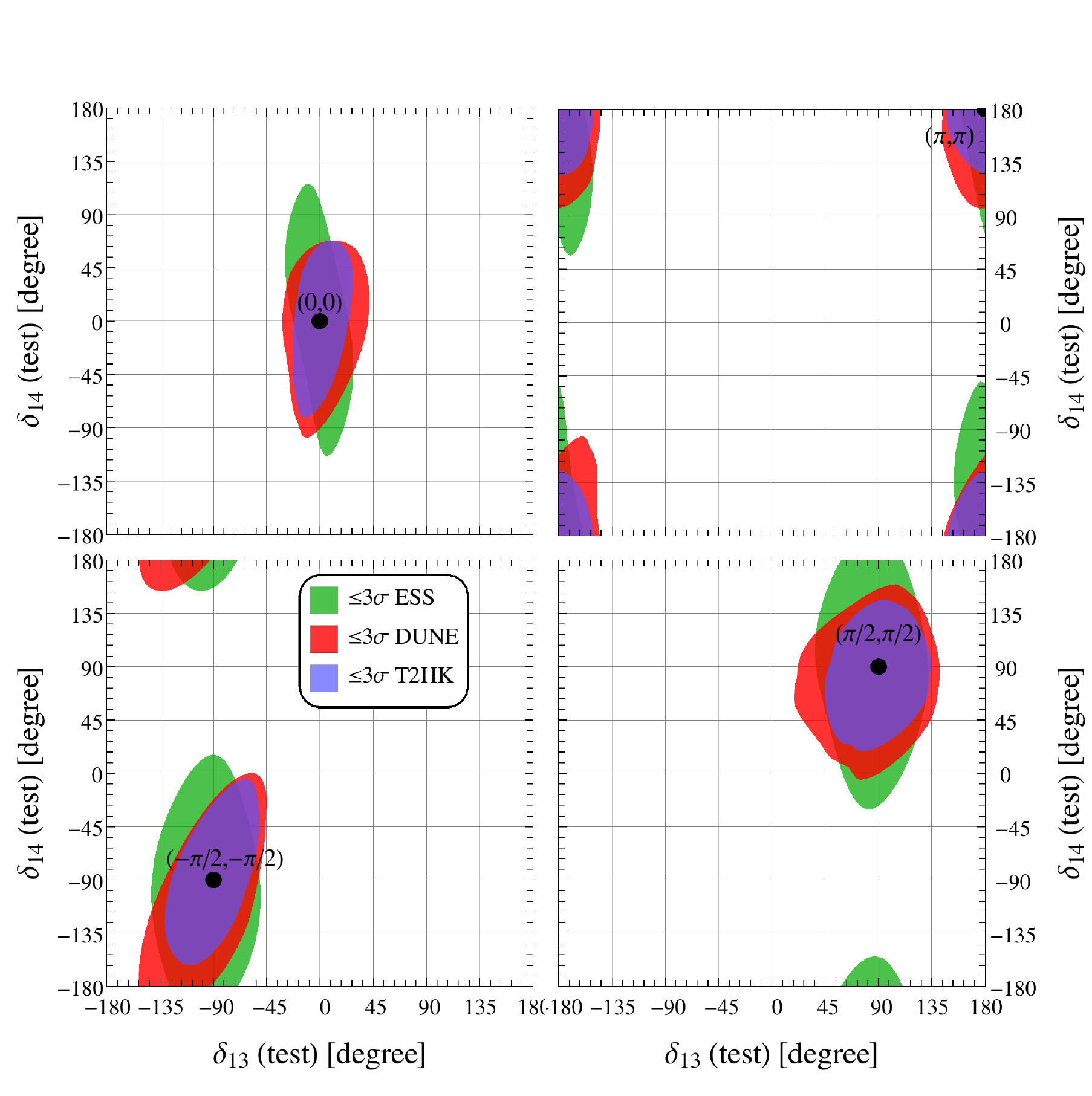}}
\vspace{-0.3cm}
\caption{Reconstructed regions for the two CP phases $\delta_{13}$ and $\delta_{14}$ 
for the four benchmark pairs of their true values indicated in each panel. Results are 
shown for the three different experimental setups: ESS$\nu$SB, DUNE, and T2HK.
We have fixed the NH as the true and test hierarchy. The contours correspond 
to 3$\sigma$ (1 d.o.f.) confidence level.}
\label{CPV_rec_2}
\end{figure}

\subsection{Reconstructing the CP phases}

So far we have discussed  the sensitivity to the CPV
induced by the two CP phases $\delta_{13}$ and 
$\delta_{14}$. Here, we study the ability of the 
ESS$\nu$SB setup to reconstruct the two CP phases.
With this aim, we focus on the four benchmark cases shown 
in Fig.~\ref{CPV_rec_1}. The first two panels correspond to the
CP-conserving cases $(0,0)$ and $(\pi,\pi)$. The lower panels 
represent two CP-violating scenarios $(-\pi/2, -\pi/2)$ and $(\pi/2, \pi/2)$.
In each panel, we show the regions reconstructed close to the true values
of the two CP phases. In this figure we have fixed the NH as the
true and test hierarchy. The contours are shown for the 
two different confidence levels: 2$\sigma$ and 3$\sigma$ (1 d.o.f.). 
The typical 1$\sigma$ level uncertainty on the reconstructed 
CP phases is approximately $15^0$ ($35^0$) for $\delta_{13}$
($\delta_{14}$)\footnote{Note that both $\delta_{13}$ and 
$\delta_{14}$ are cyclic variables, hence, the four corners 
in the top right panel of Fig.~\ref{CPV_rec_1} give rise to a 
unique connected region.}.

We end this section by comparing the performance 
of the ESS$\nu$SB setup with the two other proposed 
long-baseline facilities: DUNE%
\footnote{To simulate the DUNE setup, we consider the reference design
as mentioned in the Conceptual Design Report 
(CDR)~\cite{Acciarri:2015uup} and the necessary simulation files
for the GLoBES software are taken from~\cite{Alion:2016uaj}.} 
and 
T2HK\footnote{To estimate the reconstruction capability 
of the T2HK setup, we closely follow the experimental 
configurations as described in Ref.~\cite{Abe:2014oxa,Abe:2015zbg}.}.
In Fig.~\ref{CPV_rec_2}, we show the reconstructed 
regions for ESS$\nu$SB, DUNE, and T2HK for the same
benchmark values of the true phases considered in 
Fig.~\ref{CPV_rec_1}. To have the visual clearness,
we only depict the 3$\sigma$ (1 d.o.f.) contours.
While making this plot, we consider the NH as the true
and test hierarchy. The performance of ESS$\nu$SB
in reconstructing $\delta_{13}$ is almost similar to that of DUNE and T2HK.
In contrast, the reconstruction of $\delta_{14}$ is slightly better for
T2HK and DUNE as compared to ESS$\nu$SB.

\section{Conclusions and Outlook}
\label{Conclusions}

We have studied in detail the potential of ESS$\nu$SB in the presence of
a light eV-scale sterile neutrino with an emphasis on the CPV searches.
We have presented our results assuming a baseline of 540 km, which
provides a platform to exploit the featutres of the second oscillation maximum.
We have found that the sensitivity to CPV driven
by the standard CP phase $\delta_{13}$ substantially deteriorates
with respect to the standard 3-flavor case.
More specifically, the maximal sensitivity (assumed for $\delta_{13}$ $\sim$ $\pm$ $90^0$) 
drops from  $8\sigma$ down to $4.5\sigma$ if the size of the mixing angles 
$\theta_{14}$ and $\theta_{24}$ is similar to that of $\theta_{13}$. 
The sensitivity to the CPV induced by $\delta_{14}$ is modest and never
exceeds the 2$\sigma$ level for the baseline choice of 540 km. We have also studied the ability of reconstructing 
the two phases $\delta_{13}$ and $\delta_{14}$. 
The 1$\sigma$ error on $\delta_{13}$ ($\delta_{14}$) 
is $\sim15^0$ ($35^0$). As far as the octant of $\theta_{23}$ is concerned,
the benchmark setup under consideration for the ESS$\nu$SB experiment
provides poor results in the 3-flavor scenario and performs even worse in 3+1 scheme.
Needless to mention that ESS$\nu$SB benefits a lot from working at the second oscillation maximum
and provides excellent sensitivity to CPV in 3$\nu$ scheme. However, in the present work
we find that  this setup with a baseline of 540 km is not optimal for exploring fundamental neutrino properties in 3+1 scenario.

\subsubsection*{Acknowledgments}

S.K.A. is supported by the DST/INSPIRE Research Grant [IFA-PH-12] from
the Department of Science and Technology (DST), India and the Indian 
National Science Academy (INSA) Young Scientist Project 
[INSA/SP/YSP/144/2017/1578]. S.K.A. would like to thank Tord Ekelof for useful discussions.
S.S.C. acknowledges the partial support from the DST/INSPIRE Research Grant 
[IFA-PH-12], Department of Science and Technology, India during the initial stage 
of this project at Institute of Physics (IOP), Bhubaneswar, India. This work made 
extensive use of the cluster facilities at IOP, Bhubaneswar, India.
A.P. acknowledges partial support by the research grant number 
2017W4HA7S ``NAT-NET: Neutrino and Astroparticle Theory Network'' 
under the program PRIN 2017 funded by the 
Italian Ministero dell'Istruzione, dell'Universit\`a e della Ricerca (MIUR) 
and by the research project {\em TAsP} funded 
by the Instituto Nazionale di Fisica Nucleare (INFN).

\bibliographystyle{JHEP}
\bibliography{Sterile-References}

\providecommand{\href}[2]{#2}\begingroup\raggedright\begin{thebibliography}{10}

\bibitem{Abazajian:2012ys}
K.~N. Abazajian et~al., {\it {Light Sterile Neutrinos: A White Paper}},
  \href{http://arxiv.org/abs/1204.5379}{{\tt arXiv:1204.5379}}.

\bibitem{Palazzo:2013me}
A.~Palazzo, {\it {Phenomenology of light sterile neutrinos: a brief review}},
  {\em Mod. Phys. Lett.} {\bf A28} (2013) 1330004,
  [\href{http://arxiv.org/abs/1302.1102}{{\tt arXiv:1302.1102}}].

\bibitem{Gariazzo:2015rra}
S.~Gariazzo, C.~Giunti, M.~Laveder, Y.~F. Li, and E.~M. Zavanin, {\it {Light
  sterile neutrinos}},  \href{http://arxiv.org/abs/1507.08204}{{\tt
  arXiv:1507.08204}}.

\bibitem{Giunti:2015wnd}
C.~Giunti, {\it {Light Sterile Neutrinos: Status and Perspectives}},  {\em
  Nucl. Phys.} {\bf B908} (2016) 336--353,
  [\href{http://arxiv.org/abs/1512.04758}{{\tt arXiv:1512.04758}}].

\bibitem{Giunti:2019aiy}
C.~Giunti and T.~Lasserre, {\it {eV-scale Sterile Neutrinos}},
  \href{http://arxiv.org/abs/1901.08330}{{\tt arXiv:1901.08330}}.

\bibitem{Boser:2019rta}
S.~Böser, C.~Buck, C.~Giunti, J.~Lesgourgues, L.~Ludhova, S.~Mertens,
  A.~Schukraft, and M.~Wurm, {\it {Status of Light Sterile Neutrino Searches}},
   \href{http://arxiv.org/abs/1906.01739}{{\tt arXiv:1906.01739}}.

\bibitem{Aguilar:2001ty}
{\bf LSND} Collaboration, A.~Aguilar-Arevalo et~al., {\it {Evidence for
  neutrino oscillations from the observation of anti-neutrino(electron)
  appearance in a anti-neutrino(muon) beam}},  {\em Phys. Rev.} {\bf D64}
  (2001) 112007, [\href{http://arxiv.org/abs/hep-ex/0104049}{{\tt
  hep-ex/0104049}}].

\bibitem{Aguilar-Arevalo:2018gpe}
{\bf MiniBooNE} Collaboration, A.~A. Aguilar-Arevalo et~al., {\it {Significant
  Excess of ElectronLike Events in the MiniBooNE Short-Baseline Neutrino
  Experiment}},  {\em Phys. Rev. Lett.} {\bf 121} (2018), no.~22 221801,
  [\href{http://arxiv.org/abs/1805.12028}{{\tt arXiv:1805.12028}}].

\bibitem{Mention:2011rk}
G.~Mention, M.~Fechner, T.~Lasserre, T.~Mueller, D.~Lhuillier, et~al., {\it
  {The Reactor Antineutrino Anomaly}},  {\em Phys.Rev.} {\bf D83} (2011)
  073006, [\href{http://arxiv.org/abs/1101.2755}{{\tt arXiv:1101.2755}}].

\bibitem{Hampel:1997fc}
{\bf GALLEX} Collaboration, W.~Hampel et~al., {\it {Final results of the Cr-51
  neutrino source experiments in GALLEX}},  {\em Phys. Lett.} {\bf B420} (1998)
  114--126.

\bibitem{Abdurashitov:2005tb}
J.~N. Abdurashitov et~al., {\it {Measurement of the response of a Ga solar
  neutrino experiment to neutrinos from an Ar-37 source}},  {\em Phys. Rev.}
  {\bf C73} (2006) 045805, [\href{http://arxiv.org/abs/nucl-ex/0512041}{{\tt
  nucl-ex/0512041}}].

\bibitem{MINOS:2016viw}
{\bf MINOS} Collaboration, P.~Adamson et~al., {\it {Search for Sterile
  Neutrinos Mixing with Muon Neutrinos in MINOS}},  {\em Phys. Rev. Lett.} {\bf
  117} (2016), no.~15 151803, [\href{http://arxiv.org/abs/1607.01176}{{\tt
  arXiv:1607.01176}}].

\bibitem{Adamson:2017uda}
{\bf MINOS+} Collaboration, P.~Adamson et~al., {\it {Search for sterile
  neutrinos in MINOS and MINOS+ using a two-detector fit}},  {\em Phys. Rev.
  Lett.} {\bf 122} (2019), no.~9 091803,
  [\href{http://arxiv.org/abs/1710.06488}{{\tt arXiv:1710.06488}}].

\bibitem{Adamson:2017zcg}
{\bf NOvA} Collaboration, P.~Adamson et~al., {\it {Search for active-sterile
  neutrino mixing using neutral-current interactions in NOvA}},  {\em Phys.
  Rev.} {\bf D96} (2017), no.~7 072006,
  [\href{http://arxiv.org/abs/1706.04592}{{\tt arXiv:1706.04592}}].

\bibitem{Abe:2019fyx}
{\bf T2K} Collaboration, K.~Abe et~al., {\it {Search for light sterile
  neutrinos with the T2K far detector Super-Kamiokande at a baseline of 295
  km}},  {\em Phys. Rev.} {\bf D99} (2019), no.~7 071103,
  [\href{http://arxiv.org/abs/1902.06529}{{\tt arXiv:1902.06529}}].

\bibitem{An:2016luf}
{\bf Daya Bay} Collaboration, F.~P. An et~al., {\it {Improved Search for a
  Light Sterile Neutrino with the Full Configuration of the Daya Bay
  Experiment}},  {\em Phys. Rev. Lett.} {\bf 117} (2016), no.~15 151802,
  [\href{http://arxiv.org/abs/1607.01174}{{\tt arXiv:1607.01174}}].

\bibitem{Adamson:2016jku}
{\bf Daya Bay, MINOS} Collaboration, P.~Adamson et~al., {\it {Limits on Active
  to Sterile Neutrino Oscillations from Disappearance Searches in the MINOS,
  Daya Bay, and Bugey-3 Experiments}},  {\em Phys. Rev. Lett.} {\bf 117}
  (2016), no.~15 151801, [\href{http://arxiv.org/abs/1607.01177}{{\tt
  arXiv:1607.01177}}]. [Addendum: Phys. Rev. Lett.117,no.20,209901(2016)].

\bibitem{Danilov:2019aef}
{\bf DANSS} Collaboration, M.~Danilov, {\it {Recent results of the DANSS
  experiment}},  in {\em {2019 European Physical Society Conference on High
  Energy Physics (EPS-HEP2019) Ghent, Belgium, July 10-17, 2019}}, 2019.
\newblock \href{http://arxiv.org/abs/1911.10140}{{\tt arXiv:1911.10140}}.

\bibitem{Ko:2016owz}
{\bf NEOS} Collaboration, Y.~J. Ko et~al., {\it {Sterile Neutrino Search at the
  NEOS Experiment}},  {\em Phys. Rev. Lett.} {\bf 118} (2017), no.~12 121802,
  [\href{http://arxiv.org/abs/1610.05134}{{\tt arXiv:1610.05134}}].

\bibitem{Abe:2014gda}
{\bf Super-Kamiokande} Collaboration, K.~Abe et~al., {\it {Limits on sterile
  neutrino mixing using atmospheric neutrinos in Super-Kamiokande}},  {\em
  Phys. Rev.} {\bf D91} (2015) 052019,
  [\href{http://arxiv.org/abs/1410.2008}{{\tt arXiv:1410.2008}}].

\bibitem{Aartsen:2017bap}
{\bf IceCube} Collaboration, M.~G. Aartsen et~al., {\it {Search for sterile
  neutrino mixing using three years of IceCube DeepCore data}},  {\em Phys.
  Rev.} {\bf D95} (2017), no.~11 112002,
  [\href{http://arxiv.org/abs/1702.05160}{{\tt arXiv:1702.05160}}].

\bibitem{Albert:2018mnz}
{\bf ANTARES} Collaboration, A.~Albert et~al., {\it {Measuring the atmospheric
  neutrino oscillation parameters and constraining the 3+1 neutrino model with
  ten years of ANTARES data}},  {\em JHEP} {\bf 06} (2019) 113,
  [\href{http://arxiv.org/abs/1812.08650}{{\tt arXiv:1812.08650}}].

\bibitem{Giunti:2009xz}
C.~Giunti and Y.~F. Li, {\it {Matter Effects in Active-Sterile Solar Neutrino
  Oscillations}},  {\em Phys. Rev.} {\bf D80} (2009) 113007,
  [\href{http://arxiv.org/abs/0910.5856}{{\tt arXiv:0910.5856}}].

\bibitem{Palazzo:2011rj}
A.~Palazzo, {\it {Testing the very-short-baseline neutrino anomalies at the
  solar sector}},  {\em Phys. Rev.} {\bf D83} (2011) 113013,
  [\href{http://arxiv.org/abs/1105.1705}{{\tt arXiv:1105.1705}}].

\bibitem{Palazzo:2012yf}
A.~Palazzo, {\it {An estimate of $\theta_{14}$ independent of the reactor
  antineutrino flux determinations}},  {\em Phys. Rev.} {\bf D85} (2012)
  077301, [\href{http://arxiv.org/abs/1201.4280}{{\tt arXiv:1201.4280}}].

\bibitem{Lasserre:2014ita}
T.~Lasserre, {\it {Light Sterile Neutrinos in Particle Physics: Experimental
  Status}},  {\em Phys. Dark Univ.} {\bf 4} (2014) 81--85,
  [\href{http://arxiv.org/abs/1404.7352}{{\tt arXiv:1404.7352}}].

\bibitem{Klop:2014ima}
N.~Klop and A.~Palazzo, {\it {Imprints of CP violation induced by sterile
  neutrinos in T2K data}},  {\em Phys. Rev.} {\bf D91} (2015), no.~7 073017,
  [\href{http://arxiv.org/abs/1412.7524}{{\tt arXiv:1412.7524}}].

\bibitem{Feldman:2012qt}
G.~Feldman, J.~Hartnell, and T.~Kobayashi, {\it {Long-baseline neutrino
  oscillation experiments}},  {\em Adv.High Energy Phys.} {\bf 2013} (2013)
  475749, [\href{http://arxiv.org/abs/1210.1778}{{\tt arXiv:1210.1778}}].

\bibitem{Pascoli:2013wca}
S.~Pascoli and T.~Schwetz, {\it {Prospects for neutrino oscillation physics}},
  {\em Adv.High Energy Phys.} {\bf 2013} (2013) 503401.

\bibitem{Agarwalla:2013hma}
S.~K. Agarwalla, S.~Prakash, and S.~Uma~Sankar, {\it {Exploring the three
  flavor effects with future superbeams using liquid argon detectors}},  {\em
  JHEP} {\bf 1403} (2014) 087, [\href{http://arxiv.org/abs/1304.3251}{{\tt
  arXiv:1304.3251}}].

\bibitem{Agarwalla:2014fva}
S.~K. Agarwalla, {\it {Physics Potential of Long-Baseline Experiments}},  {\em
  Adv.High Energy Phys.} {\bf 2014} (2014) 457803,
  [\href{http://arxiv.org/abs/1401.4705}{{\tt arXiv:1401.4705}}].

\bibitem{Abe:2015zbg}
{\bf Hyper-Kamiokande Proto-Collaboration} Collaboration, K.~Abe et~al., {\it
  {Physics potential of a long-baseline neutrino oscillation experiment using a
  J-PARC neutrino beam and Hyper-Kamiokande}},  {\em PTEP} {\bf 2015} (2015)
  053C02, [\href{http://arxiv.org/abs/1502.05199}{{\tt arXiv:1502.05199}}].

\bibitem{Stanco:2015ejj}
L.~Stanco, {\it {A View of Neutrino Studies with the Next Generation
  Facilities}},  {\em Rev. Phys.} {\bf 1} (2016) 90--100,
  [\href{http://arxiv.org/abs/1511.09409}{{\tt arXiv:1511.09409}}].

\bibitem{Acciarri:2015uup}
{\bf DUNE} Collaboration, R.~Acciarri et~al., {\it {Long-Baseline Neutrino
  Facility (LBNF) and Deep Underground Neutrino Experiment (DUNE)}},
  \href{http://arxiv.org/abs/1512.06148}{{\tt arXiv:1512.06148}}.

\bibitem{Abe:2016ero}
{\bf Hyper-Kamiokande} Collaboration, K.~Abe et~al., {\it {Physics potentials
  with the second Hyper-Kamiokande detector in Korea}},  {\em PTEP} {\bf 2018}
  (2018), no.~6 063C01, [\href{http://arxiv.org/abs/1611.06118}{{\tt
  arXiv:1611.06118}}].

\bibitem{Agarwalla:2017nld}
S.~K. Agarwalla, M.~Ghosh, and S.~K. Raut, {\it {A hybrid setup for fundamental
  unknowns in neutrino oscillations using T2HK ($\nu$) and $\mu$-DAR
  ($\bar{\nu}$)}},  {\em JHEP} {\bf 05} (2017) 115,
  [\href{http://arxiv.org/abs/1704.06116}{{\tt arXiv:1704.06116}}].

\bibitem{Abi:2018dnh}
{\bf DUNE} Collaboration, B.~Abi et~al., {\it {The DUNE Far Detector Interim
  Design Report Volume 1: Physics, Technology and Strategies}},
  \href{http://arxiv.org/abs/1807.10334}{{\tt arXiv:1807.10334}}.

\bibitem{Hollander:2014iha}
D.~Hollander and I.~Mocioiu, {\it {Minimal 3+2 sterile neutrino model at
  LBNE}},  {\em Phys. Rev.} {\bf D91} (2015), no.~1 013002,
  [\href{http://arxiv.org/abs/1408.1749}{{\tt arXiv:1408.1749}}].

\bibitem{Berryman:2015nua}
J.~M. Berryman, A.~de~Gouv{\^e}a, K.~J. Kelly, and A.~Kobach, {\it {Sterile
  neutrino at the Deep Underground Neutrino Experiment}},  {\em Phys. Rev.}
  {\bf D92} (2015), no.~7 073012, [\href{http://arxiv.org/abs/1507.03986}{{\tt
  arXiv:1507.03986}}].

\bibitem{Gandhi:2015xza}
R.~Gandhi, B.~Kayser, M.~Masud, and S.~Prakash, {\it {The impact of sterile
  neutrinos on CP measurements at long baselines}},  {\em JHEP} {\bf 11} (2015)
  039, [\href{http://arxiv.org/abs/1508.06275}{{\tt arXiv:1508.06275}}].

\bibitem{Agarwalla:2016xxa}
S.~K. Agarwalla, S.~S. Chatterjee, and A.~Palazzo, {\it {Physics Reach of DUNE
  with a Light Sterile Neutrino}},  {\em JHEP} {\bf 09} (2016) 016,
  [\href{http://arxiv.org/abs/1603.03759}{{\tt arXiv:1603.03759}}].

\bibitem{Agarwalla:2016xlg}
S.~K. Agarwalla, S.~S. Chatterjee, and A.~Palazzo, {\it {Octant of
  $\theta_{23}$ in danger with a light sterile neutrino}},  {\em Phys. Rev.
  Lett.} {\bf 118} (2017), no.~3 031804,
  [\href{http://arxiv.org/abs/1605.04299}{{\tt arXiv:1605.04299}}].

\bibitem{Coloma:2017ptb}
P.~Coloma, D.~V. Forero, and S.~J. Parke, {\it {DUNE Sensitivities to the
  Mixing between Sterile and Tau Neutrinos}},  {\em JHEP} {\bf 07} (2018) 079,
  [\href{http://arxiv.org/abs/1707.05348}{{\tt arXiv:1707.05348}}].

\bibitem{Choubey:2017cba}
S.~Choubey, D.~Dutta, and D.~Pramanik, {\it {Imprints of a light Sterile
  Neutrino at DUNE, T2HK and T2HKK}},
  \href{http://arxiv.org/abs/1704.07269}{{\tt arXiv:1704.07269}}.

\bibitem{Choubey:2017ppj}
S.~Choubey, D.~Dutta, and D.~Pramanik, {\it {Measuring the Sterile Neutrino CP
  Phase at DUNE and T2HK}},  \href{http://arxiv.org/abs/1711.07464}{{\tt
  arXiv:1711.07464}}.

\bibitem{Agarwalla:2018t2hk}
S.~K. Agarwalla, S.~S. Chatterjee, and A.~Palazzo, {\it {Signatures of a Light
  Sterile Neutrino in T2HK}},  {\em JHEP} {\bf 04} (2018) 091,
  [\href{http://arxiv.org/abs/1801.04855}{{\tt arXiv:1801.04855}}].

\bibitem{Haba:2018klh}
N.~Haba, Y.~Mimura, and T.~Yamada, {\it {On $\theta_{23}$ Octant Measurement in
  $3+1$ Neutrino Oscillations in T2HKK}},
  \href{http://arxiv.org/abs/1812.10940}{{\tt arXiv:1812.10940}}.

\bibitem{Donini:2001xy}
A.~Donini and D.~Meloni, {\it {The 2+2 and 3+1 four family neutrino mixing at
  the neutrino factory}},  {\em Eur. Phys. J.} {\bf C22} (2001) 179--186,
  [\href{http://arxiv.org/abs/hep-ph/0105089}{{\tt hep-ph/0105089}}].

\bibitem{Donini:2001xp}
A.~Donini, M.~Lusignoli, and D.~Meloni, {\it {Telling three neutrinos from four
  neutrinos at the neutrino factory}},  {\em Nucl. Phys.} {\bf B624} (2002)
  405--422, [\href{http://arxiv.org/abs/hep-ph/0107231}{{\tt hep-ph/0107231}}].

\bibitem{Donini:2007yf}
A.~Donini, M.~Maltoni, D.~Meloni, P.~Migliozzi, and F.~Terranova, {\it {3+1
  sterile neutrinos at the CNGS}},  {\em JHEP} {\bf 12} (2007) 013,
  [\href{http://arxiv.org/abs/0704.0388}{{\tt arXiv:0704.0388}}].

\bibitem{Dighe:2007uf}
A.~Dighe and S.~Ray, {\it {Signatures of heavy sterile neutrinos at long
  baseline experiments}},  {\em Phys. Rev.} {\bf D76} (2007) 113001,
  [\href{http://arxiv.org/abs/0709.0383}{{\tt arXiv:0709.0383}}].

\bibitem{Donini:2008wz}
A.~Donini, K.-i. Fuki, J.~Lopez-Pavon, D.~Meloni, and O.~Yasuda, {\it {The
  Discovery channel at the Neutrino Factory: numu to nutau pointing to sterile
  neutrinos}},  {\em JHEP} {\bf 08} (2009) 041,
  [\href{http://arxiv.org/abs/0812.3703}{{\tt arXiv:0812.3703}}].

\bibitem{Yasuda:2010rj}
O.~Yasuda, {\it {Sensitivity to sterile neutrino mixings and the discovery
  channel at a neutrino factory}},  in {\em {Physics beyond the standard models
  of particles, cosmology and astrophysics. Proceedings, 5th International
  Conference, Beyond 2010, Cape Town, South Africa, February 1-6, 2010}},
  pp.~300--313, 2011.
\newblock \href{http://arxiv.org/abs/1004.2388}{{\tt arXiv:1004.2388}}.

\bibitem{Meloni:2010zr}
D.~Meloni, J.~Tang, and W.~Winter, {\it {Sterile neutrinos beyond LSND at the
  Neutrino Factory}},  {\em Phys. Rev.} {\bf D82} (2010) 093008,
  [\href{http://arxiv.org/abs/1007.2419}{{\tt arXiv:1007.2419}}].

\bibitem{Bhattacharya:2011ee}
B.~Bhattacharya, A.~M. Thalapillil, and C.~E.~M. Wagner, {\it {Implications of
  sterile neutrinos for medium/long-baseline neutrino experiments and the
  determination of $\theta_{13}$}},  {\em Phys. Rev.} {\bf D85} (2012) 073004,
  [\href{http://arxiv.org/abs/1111.4225}{{\tt arXiv:1111.4225}}].

\bibitem{Donini:2012tt}
A.~Donini, P.~Hernandez, J.~Lopez-Pavon, M.~Maltoni, and T.~Schwetz, {\it {The
  minimal 3+2 neutrino model versus oscillation anomalies}},  {\em JHEP} {\bf
  07} (2012) 161, [\href{http://arxiv.org/abs/1205.5230}{{\tt
  arXiv:1205.5230}}].

\bibitem{Gandhi:2017vzo}
R.~Gandhi, B.~Kayser, S.~Prakash, and S.~Roy, {\it {What measurements of
  neutrino neutral current events can reveal}},  {\em JHEP} {\bf 11} (2017)
  202, [\href{http://arxiv.org/abs/1708.01816}{{\tt arXiv:1708.01816}}].

\bibitem{deSalas:2017kay}
P.~F. de~Salas, D.~V. Forero, C.~A. Ternes, M.~Tortola, and J.~W.~F. Valle,
  {\it {Status of neutrino oscillations 2017}},
  \href{http://arxiv.org/abs/1708.01186}{{\tt arXiv:1708.01186}}.

\bibitem{Capozzi:2018ubv}
F.~Capozzi, E.~Lisi, A.~Marrone, and A.~Palazzo, {\it {Current unknowns in the
  three neutrino framework}},  {\em Prog. Part. Nucl. Phys.} {\bf 102} (2018)
  48--72, [\href{http://arxiv.org/abs/1804.09678}{{\tt arXiv:1804.09678}}].

\bibitem{Esteban:2018azc}
I.~Esteban, M.~C. Gonzalez-Garcia, A.~Hernandez-Cabezudo, M.~Maltoni, and
  T.~Schwetz, {\it {Global analysis of three-flavour neutrino oscillations:
  synergies and tensions in the determination of $\theta_23, \delta_CP$, and
  the mass ordering}},  {\em JHEP} {\bf 01} (2019) 106,
  [\href{http://arxiv.org/abs/1811.05487}{{\tt arXiv:1811.05487}}].

\bibitem{Capozzi:2016vac}
F.~Capozzi, C.~Giunti, M.~Laveder, and A.~Palazzo, {\it {Joint short- and
  long-baseline constraints on light sterile neutrinos}},  {\em Phys. Rev.}
  {\bf D95} (2017), no.~3 033006, [\href{http://arxiv.org/abs/1612.07764}{{\tt
  arXiv:1612.07764}}].

\bibitem{Gariazzo:2017fdh}
S.~Gariazzo, C.~Giunti, M.~Laveder, and Y.~F. Li, {\it {Updated Global 3+1
  Analysis of Short-BaseLine Neutrino Oscillations}},  {\em JHEP} {\bf 06}
  (2017) 135, [\href{http://arxiv.org/abs/1703.00860}{{\tt arXiv:1703.00860}}].

\bibitem{Dentler:2018sju}
M.~Dentler, A.~Hernandez-Cabezudo, J.~Kopp, P.~A.~N. Machado, M.~Maltoni,
  I.~Martinez-Soler, and T.~Schwetz, {\it {Updated Global Analysis of Neutrino
  Oscillations in the Presence of eV-Scale Sterile Neutrinos}},  {\em JHEP}
  {\bf 08} (2018) 010, [\href{http://arxiv.org/abs/1803.10661}{{\tt
  arXiv:1803.10661}}].

\bibitem{Diaz:2019fwt}
A.~Diaz, C.~A. Arg{\~A}¼elles, G.~H. Collin, J.~M. Conrad, and M.~H. Shaevitz,
  {\it {Where Are We With Light Sterile Neutrinos?}},
  \href{http://arxiv.org/abs/1906.00045}{{\tt arXiv:1906.00045}}.

\bibitem{Baussan:2013zcy}
{\bf ESS$\nu$SB} Collaboration, E.~Baussan et~al., {\it {A very intense
  neutrino super beam experiment for leptonic CP violation discovery based on
  the European spallation source linac}},  {\em Nucl. Phys.} {\bf B885} (2014)
  127--149, [\href{http://arxiv.org/abs/1309.7022}{{\tt arXiv:1309.7022}}].

\bibitem{Dracos:2016wso}
{\bf ESS$\nu$SB project} Collaboration, M.~Dracos, {\it {The ESS$\nu$SB Project
  for Leptonic CP Violation Discovery based on the European Spallation Source
  Linac}},  {\em Nucl. Part. Phys. Proc.} {\bf 273-275} (2016) 1726--1731.

\bibitem{Dracos:2018jsn}
M.~Dracos, {\it {The European Spallation Source neutrino Super Beam}},  in {\em
  {Proceedings, Prospects in Neutrino Physics (NuPhys2017): London, UK,
  December 20-22, 2017}}, pp.~33--41, 2018.
\newblock \href{http://arxiv.org/abs/1803.10948}{{\tt arXiv:1803.10948}}.

\bibitem{Dracos:2018syh}
{\bf ESSvSB} Collaboration, M.~Dracos and T.~Ekelof, {\it {Neutrino CP
  Violation with the ESS neutrino Super Beam (ESS$\nu$SB)}},  {\em PoS} {\bf
  ICHEP2018} (2019) 524.

\bibitem{enrique}
{Enrique Fernandez Martinez}. {private communication}, {2013}.

\bibitem{Agostino:2012fd}
{\bf MEMPHYS Collaboration} Collaboration, L.~Agostino et~al., {\it {Study of
  the performance of a large scale water-Cherenkov detector (MEMPHYS)}},  {\em
  JCAP} {\bf 1301} (2013) 024, [\href{http://arxiv.org/abs/1206.6665}{{\tt
  arXiv:1206.6665}}].

\bibitem{luca}
{Luca Agostino}. {private communication}, {2013}.

\bibitem{Agarwalla:2014tpa}
S.~K. Agarwalla, S.~Choubey, and S.~Prakash, {\it {Probing Neutrino Oscillation
  Parameters using High Power Superbeam from ESS}},  {\em JHEP} {\bf 12} (2014)
  020, [\href{http://arxiv.org/abs/1406.2219}{{\tt arXiv:1406.2219}}].

\bibitem{Huber:2004ka}
P.~Huber, M.~Lindner, and W.~Winter, {\it {Simulation of long-baseline neutrino
  oscillation experiments with GLoBES (General Long Baseline Experiment
  Simulator)}},  {\em Comput.Phys.Commun.} {\bf 167} (2005) 195,
  [\href{http://arxiv.org/abs/hep-ph/0407333}{{\tt hep-ph/0407333}}].

\bibitem{Huber:2007ji}
P.~Huber, J.~Kopp, M.~Lindner, M.~Rolinec, and W.~Winter, {\it {New features in
  the simulation of neutrino oscillation experiments with GLoBES 3.0: General
  Long Baseline Experiment Simulator}},  {\em Comput.Phys.Commun.} {\bf 177}
  (2007) 432--438, [\href{http://arxiv.org/abs/hep-ph/0701187}{{\tt
  hep-ph/0701187}}].

\bibitem{Kopp:NSI}
J.~Kopp, {\it {Sterile neutrinos and non-standard neutrino interactions in
  GLoBES}},  {\em
  https://www.mpi-hd.mpg.de/personalhomes/globes/tools/snu-1.0.pdf} (2010).

\bibitem{Agarwalla:2016mrc}
S.~K. Agarwalla, S.~S. Chatterjee, A.~Dasgupta, and A.~Palazzo, {\it {Discovery
  Potential of T2K and NOvA in the Presence of a Light Sterile Neutrino}},
  {\em JHEP} {\bf 02} (2016) 111, [\href{http://arxiv.org/abs/1601.05995}{{\tt
  arXiv:1601.05995}}].

\bibitem{PREM:1981}
A.~M. Dziewonski and D.~L. Anderson, {\it Preliminary reference earth model},
  {\em Physics of the Earth and Planetary Interiors} {\bf 25} (1981) 297--356.

\bibitem{Huber:2002mx}
P.~Huber, M.~Lindner, and W.~Winter, {\it {Superbeams versus neutrino
  factories}},  {\em Nucl. Phys.} {\bf B645} (2002) 3--48,
  [\href{http://arxiv.org/abs/hep-ph/0204352}{{\tt hep-ph/0204352}}].

\bibitem{Fogli:2002pt}
G.~L. Fogli, E.~Lisi, A.~Marrone, D.~Montanino, and A.~Palazzo, {\it {Getting
  the most from the statistical analysis of solar neutrino oscillations}},
  {\em Phys. Rev.} {\bf D66} (2002) 053010,
  [\href{http://arxiv.org/abs/hep-ph/0206162}{{\tt hep-ph/0206162}}].

\bibitem{Alion:2016uaj}
{\bf DUNE} Collaboration, T.~Alion et~al., {\it {Experiment Simulation
  Configurations Used in DUNE CDR}},
  \href{http://arxiv.org/abs/1606.09550}{{\tt arXiv:1606.09550}}.

\bibitem{Abe:2014oxa}
{\bf Hyper-Kamiokande Working Group} Collaboration, K.~Abe et~al., {\it {A Long
  Baseline Neutrino Oscillation Experiment Using J-PARC Neutrino Beam and
  Hyper-Kamiokande}},  \href{http://arxiv.org/abs/1412.4673}{{\tt
  arXiv:1412.4673}}.

\end{thebibliography}\endgroup

\end{document}